\documentclass[aps,amsmath,amssymb,pra,twocolumn,a4paper,reprint,eprint,longbibliography,footinbib]{revtex4-1}
\pdfoutput=1
\usepackage{graphicx}
\usepackage{dcolumn}
\usepackage{bm}
\usepackage{tabularx}
\usepackage{xcolor}
\usepackage{overpic}
\usepackage{multirow}
\usepackage{hyperref}

\usepackage{tablefootnote}
\usepackage{epsfig}
\usepackage{multirow}
\usepackage{hhline}
\usepackage{pifont} 
\newcommand{\tickYes}{\color{blue}\checkmark} 
\newcommand{\tickNo}{\hspace{1pt}\color{red}\ding{55}}

\newcommand{\ignore}[1]{}

\begin{document}

\title{Bound states around impurities in a superconducting bilayer}

\author{Yufei Zhu}
\author{Nico A. Hackner}
\author{P. M. R. Brydon}
\email{philip.brydon@otago.ac.nz}
 \affiliation{Department of Physics and MacDiarmid Institute for Advanced Materials and Nanotechnology,
University of Otago, P.O. Box 56, Dunedin 9054, New Zealand}

\begin{abstract}
We theoretically study the appearance of bound states around impurities in
a superconducting bilayer. We focus our attention on $s$-wave pairing,
which includes unconventional odd-parity states permitted
by the layer degree of freedom. Utilizing numerical mean-field and
analytical $T$-matrix methods, we survey the bound state spectrum
produced by momentum-independent impurity potentials in this
model. For even-parity $s$-wave pairing, bound states are only found
for impurities which break time-reversal symmetry. For odd-parity
$s$-wave states, in contrast, bound states are generically found for
all impurity potentials, and fall into six distinct categories. This
categorization remains valid for nodal gaps.  
Our results are conveniently understood in terms of the
``superconducting fitness'' concept, and show an interplay between the
pair-breaking effects of the impurity and the normal-state band structure.
\end{abstract}

\date{\today}

\maketitle
 
\section{\label{sec:intro}Introduction}

The robustness of conventional superconductors to the presence of non-magnetic impurities is famously guaranteed by Anderson's theorem and the isotropic gap function~\cite{Anderson}. However, if the disorder breaks time-reversal symmetry or the superconductor has an unconventional sign-changing gap, introducing a finite concentration of impurities rapidly suppresses the critical temperature to zero~\cite{Mineev1999}. This pair-breaking effect is also evidenced by a \emph{single} impurity with the appearance of bound states localized at the impurity with energy within the superconducting gap. So-called Yu-Shiba-Rusinov states were first proposed for magnetic impurities in conventional superconductors~\cite{Yu1965,Shiba1968,Rusinov1969}. Impurity bound states also appear in fully-gapped unconventional superconductors for both magnetic and nonmagnetic impurities~\cite{Wang2004,Wang2012,Kim2015,Mashkoori2017}, while subgap resonances appear in nodal superconductors due to the hybridization with the nodal quasiparticles~\cite{Balatsky1995,Liu2008}.
Impurity bound states have attracted much attention over the
last several decades~\cite{BalatskyRMP2006} as they can be directly
detected by scanning 
tunneling microscopy experiments~\cite{Yazdani1997}, and
hence act as a test of the pairing symmetry~\cite{Hudson2001,Grothe2012}. More
recently, it has been claimed that
topological superconductivity is realized in the bands formed by overlapping
bound states on chains of magnetic impurities~\cite{NadjPerge2014,Pawlak2016,Schneider2021}.  

Recently it has been observed that the superconductivity of
Bi$_2$Se$_3$-based compounds is remarkably robust against disorder~\cite{Kriener2012,Smylie2017,Andersen2020}, despite
strong evidence that it realizes a nematic odd-parity state which
is not protected by Anderson's theorem~\cite{Yonezawa2019}. This has prompted interest in
the possibility that the strong spin-orbit coupling (SOC) in this compound
might be responsible for the enhanced robustness of the pairing
state~\cite{Michaeli2012,Nagai2014,Nagai2015,Andersen2020,Cavanagh2020,Cavanagh2021,Dentelski2020,Sato2020}. 
Particular attention has been directed at the unconventional
odd-parity $s$-wave states allowed by the sublattice
structure of these materials~\cite{FuBerg2010}, and it has been found that 
these states can be surprisingly robust against chemical potential disorder~\cite{Michaeli2012,Cavanagh2020,Cavanagh2021}.
A key parameter in this theory is the ``superconducting fitness''
\cite{Ramires2016,Ramires2018}, which measures the degree to which the
pairing state pairs electrons in the same band: the fitter the gap,
the higher the degree of intraband pairing, and
the weaker the suppression of the critical temperature by
disorder. The fitness concept has 
also been extended to other impurity potentials, allowing the identification
of impurities which are pair-breaking for a
given pairing state~\cite{Andersen2020,Timmons2020,Dentelski2020,Cavanagh2021}.

Thus far the study of impurities in odd-parity $s$-wave
superconductors has
mainly focused upon the effect on the critical temperature~\cite{Michaeli2012,Andersen2020,Cavanagh2020,Cavanagh2021,Dentelski2020,Sato2020}. The tuning of the
pair-breaking effect by the normal-state band structure should
nevertheless also influence the existence and structure of impurity
bound states in such superconductors. This motivates us to study the
bound states around 
an impurity in a minimal model with odd-parity $s$-wave states
to clarify the role of superconducting fitness. Specifically, we
consider impurities in a superconducting Rashba bilayer~\cite{Nakosai2012} using two
distinct methods: a numerical self-consistent mean-field theory for a tight-binding model, and an analytic non-self-consistent $T$-matrix
theory. Examining all momentum-independent impurity potentials, we
find that the bound states belong to distinct classes which depend on
the symmetry of each impurity and its fitness with respect to a given
pairing state. Bound states are still possible for fit impurities
when the pairing state is unfit with respect to parts of the normal
state Hamiltonian. Our results also hold for virtual bound states
which appear when the pairing state has nodes. 

Our paper is organized as follows: In Sec. \ref{sec:bilayer}, we
introduce the two-dimensional tight-binding bilayer model and our
approximation schemes. We summarize the concept of superconducting
fitness in Sec.~\ref{subsec:fitness}, while the self-consistent mean-field theory and the $T$-matrix approximation
are introduced in Secs.~\ref{subsec:mft} and~\ref{subsec:analytic}, respectively. In Sec. \ref{sec:boundstates}, we present our numerical and
analytical results for the bound states and develop the classification
scheme for the isotropic even-parity singlet state (Sec.~\ref{sub:intralayer_singlet}) and
the fully-gapped odd-parity $s$-wave states (Sec.~\ref{sub:interlayer_triplet}). In Sec. \ref{sec:virtual
  bound states}  we  
extend our analysis to nodal gaps in a three-dimensional continuum
model, and find that our symmetry classification remains valid for the
virtual bound states. In Sec. \ref{sec:discussion} we summarize our conclusions
and discuss future work.

\section{\label{sec:bilayer}Model and methods}

As a minimal model we consider superconductivity in a two-dimensional
square bilayer lattice with an impurity. This is described by the Hamiltonian
\begin{equation}\label{eq:hamiltonian}
    H = H_0  + H_{\text{imp}} + H_{\text{int}} \,.
\end{equation}
The first term describes the motion of the noninteracting electrons
\begin{equation} \label{eq:H0}
  H_0 = \sum_{\mathbf{r_i,r_j}} \mathbf{c}_\mathbf{r_i}^\dagger \mathcal{H_\mathbf{r_i,r_j}} \mathbf{c_\mathbf{r_j}} \,,
\end{equation}
where $\mathbf{c_\mathbf{r_j}} = (c_{A, \mathbf{r_j}, \uparrow}, c_{A, \mathbf{r_j}, \downarrow}, c_{B, \mathbf{r_j}, \uparrow}  c_{B, \mathbf{r_j}, \downarrow})^T$, and $c_{\eta,\mathbf{r_j},\sigma}$ is the destruction operator for a spin-$\sigma$ electron on layer $\eta=A$, $B$ at $\mathbf{r_j}$, and the matrix elements are
\begin{eqnarray} \label{eq:minbilayer}
\hspace{-0.3cm}
  \mathcal{H_\mathbf{r_i,r_j}} &=& \left[-t(\delta_{\mathbf{r_i,r_j}\pm a\hat{\bf x}}+\delta_{\mathbf{r_i,r_j} \pm a\hat{\bf y}}) - \mu \delta_{\mathbf{r_i,r_j}}\right] \eta_0 \sigma_0 \nonumber \\ &&
  + t_{\perp} \delta_{\mathbf{r_i,r_j}} \eta_x \sigma_0 + \lambda_{x,\mathbf{r_i,r_j}} \eta_z \sigma_x + \lambda_{y,\mathbf{r_i,r_j}} \eta_z \sigma_y \,,
\end{eqnarray}
where the $\sigma_\mu$ and $\eta_\mu$ are the Pauli matrices for spin
and layer degrees of freedom, respectively, and $\eta_\nu\sigma_\mu$
is to be understood as a Kronecker product. The first line in
Eq.~\ref{eq:minbilayer} includes the intralayer nearest-neighbour
hopping $t$ and the chemical potential $\mu$,
the second line describes the coupling of the layers via the hopping
$t_\perp$ between the $A$ and $B$ sites in each unit cell, and a Rashba SOC of strength $\alpha$ in each layer 
\begin{eqnarray}
    \lambda_{x,\mathbf{r_i,r_j}} &= & \pm \frac{\alpha}{2 i}\delta_{\mathbf{r_i,r_j}\pm a\hat{\bf y}} \,, \\ 
  \lambda_{y,\mathbf{r_i,r_j}} & = & \mp \frac{\alpha}{2 i}\delta_{\mathbf{r_i,r_j}\pm a\hat{\bf x}} \,.
\end{eqnarray}
The Rashba SOC is generically present as the $A$ and $B$ sites are not
inversion centres~\cite{Nakosai2012}. To preserve the global inversion symmetry, which
swaps the $A$ and $B$ layers, the Rashba SOC has an opposite sign in each layer as encoded by the $\eta_z$ Pauli matrix in Eq.~\ref{eq:minbilayer}.

The second term in Eq.~\ref{eq:hamiltonian} is
\begin{equation}
  H_{\text{imp}} = {\bf c}_{{\bf r_0}}^\dagger \hat{V}_{\text{imp}} {\bf c}_{{\bf r_0}} \,.
\end{equation}
This describes an impurity located at site ${\bf r_0}$ which couples to the electrons via one of the sixteen impurity potentials tabulated in Table~\ref{tab:imp_survey}, i.e. $\hat{V}_{\text{imp}}= V^{\alpha\beta}\eta_\alpha  \sigma_\beta$. We consider each impurity potential individually, i.e. we only take one of the $V^{\alpha\beta}=V$ to be nonzero at a time.

The layer degree of freedom allows for unconventional $s$-wave (i.e. intra-unit cell) paring states as listed in Table~\ref{tab:pairing_states}. We include these states in Eq.~\ref{eq:hamiltonian} via the phenomenological pairing interaction
\begin{eqnarray}
\label{eq:H_int}
H_\text{int} = \frac{g}{4}\sum_{\mathbf{r_j}} \left\{ \left( \mathbf{c}^\dagger_{\mathbf{r_j}} \hat{\Delta}_\nu \mathbf{c}^\dagger_{\mathbf{r_j}} \right) \left( \mathbf{ c_{r_j}} \hat{\Delta}_\nu \mathbf{c_{r_j}}\right) \right\},
\end{eqnarray}
where $g$ is the effective coupling constant and $\hat{\Delta}_\nu$ is
one of $s$-wave pairing potentials. In the following we set the coupling constant to be nonzero in a single pairing channel at a time.

Performing a mean-field decoupling of the interaction Hamiltonian in
the Cooper channel, we obtain the Bogoliubov-de Gennes (BdG) Hamiltonian \begin{equation}\label{eq:BdG}
H_{\text{BdG}} = \frac{1}{2}\sum_{\mathbf{r_i}, \mathbf{r_j}} \mathbf{\Psi}_{\mathbf{r_i}}^\dagger     \begin{pmatrix}
         \mathcal{\tilde{H}_\mathbf{r_i,r_j}} & \hat{\Delta}_\mathbf{r_i,r_j} \\
        \hat{\Delta}^\dagger_\mathbf{r_i,r_j} & -\mathcal{\tilde{H}}^T_\mathbf{r_i,r_j}
    \end{pmatrix} \mathbf{\Psi_{r_j}}\,.
\end{equation} 
Here $\mathbf{\Psi_{r_j}} = (\mathbf{c}^T_{\mathbf{r_j}},\mathbf{c}^\dagger_{\mathbf{r_j}})^T$ is an eight-component spinor of creation and annihilation operators, $\mathcal{\tilde{H}_\mathbf{r_i,r_j}} = \mathcal{H_\mathbf{r_i,r_j}} + \hat{V}_{\text{imp}}\delta_{\mathbf{r_i,r_0}}\delta_{\mathbf{r_j,r_0}}$, and the pairing term is given by 
\begin{equation}\label{eq:selfconsistent}
    \hat{\Delta}_\mathbf{r_i,r_j} = \frac{1}{2}\delta_\mathbf{r_i,r_j} g\langle \mathbf{ c_{r_j}} \hat{\Delta}_\nu \mathbf{c_{r_j}} \rangle \hat{\Delta}_\nu ,
\end{equation} 
where $\langle ... \rangle$ represents the thermally-weighted expectation value.

\subsection{Superconducting fitness}~\label{subsec:fitness}

Due to the presence of the layer degree of freedom, the normal-state Hamiltonian Eq.~\ref{eq:H0} describes a two-band electronic system. Expressed in momentum space, the band energies are given by
\begin{eqnarray} \label{eq:eigvals}
     E_{\pm,\mathbf{k}} &=& -2t\left(\cos(k_xa)+\cos(k_ya)\right)-\mu \nonumber \\
     && \pm \sqrt{ \alpha^2\text{sin}^2 ({k_x a}) + \alpha^2\text{sin}^2 ({k_y a})   + t_\perp^2} \,,
\end{eqnarray}
where $a$ is the lattice constant. The multiband nature has an important consequence for the superconductivity: Pairing can occur between electrons in both the same and different bands. Of the pairing potentials listed in Tab.~\ref{tab:pairing_states}, only the uniform singlet state pairs electrons in time-reversed states, which guarantees purely intraband pairing; the other superconducting states will in general involve both intra- and interband pairing. 

The concept of the superconducting fitness has recently been developed
as a way to quantify the degree of interband
pairing~\cite{Ramires2016,Ramires2018}. For the pairing potential
$\hat{\Delta} = \Delta_0\hat{\Delta}_\nu$, where $\Delta_0$ is the
amplitude and $\hat{\Delta}_\nu$ is a dimensionless matrix for pairing
in channel $\nu$, we define the fitness function
\begin{equation}
  F_{\nu,{\bm k}} = 
  {\cal H}_{0,{\bm k}}\hat{\Delta}_\nu - \hat{\Delta}_\nu{\cal H}^T_{0,-{\bm k}} \,, \label{eq:fitness}
\end{equation}
which is only nonzero if the interband pairing is present. 
For the two-band system considered here the fitness can be related to the gap opened at the Fermi energy 
\begin{equation}
  \Delta_{\text{eff},{\bm k}} = |\Delta_0|\sqrt{1 - \tilde{F}_{\nu,\bm{k}}} \,, \label{eq:effgap}
\end{equation}
where 
\begin{equation}
  \tilde{F}_{\nu,\bm k} = \frac{\text{Tr}\{|F_{\nu,{\bm k}}|^2\}}{(E_{+,{\bm k}} - E_{-,{\bm k}})^2}\,. \label{eq:fitnessnorm}
\end{equation}
Eq.~\ref{eq:fitnessnorm} takes values between $0$ and $1$, with $\tilde{F}_{\nu,{\bm k}} = 0$ indicating a perfectly ``fit'' pairing state where there is only intraband pairing and thus the gap is maximal, whereas $\tilde{F}_{\nu,{\bm k}}=1$ implies that only interband pairing is present and no gap is opened at the Fermi energy at weak coupling.

The concept of superconducting fitness has been extended to the impurity potential~\cite{Cavanagh2021}. In a two-band system, whether or not the impurity is pair-breaking in the channel $\nu$ is given by the solution of
\begin{equation}
  \hat{V}_{\text{imp}}\hat{\Delta}_{\nu} - \lambda \hat{\Delta}_{\nu}\hat{V}_{\text{imp}}^T = 0 \label{eq:genAnderson} \,,
\end{equation}
where $\lambda=-1$ ($1$) implies that the impurity is (is not)
pair-breaking. In the case of on-site singlet pairing this is
equivalent to whether or not the impurity breaks time-reversal
symmetry, and is thus a restatement of Anderson's
theorem~\cite{Anderson}; for the nontrivial $s$-wave pairing
potentials Eq.~\ref{eq:genAnderson} therefore establishes a generalized
Anderson's theorem. An important difference compared to the usual
Anderson's theorem is that the critical temperature of the odd-parity 
$s$-wave states is generally suppressed even when the impurity is not
pair-breaking if the fitness $0<\tilde{F}_{\nu,\bm k}<1$, albeit more
slowly compared to the usual Gor'kov theory~\cite{Mineev1999}. 

\subsection{Real-space mean-field theory}~\label{subsec:mft}

We solve the self-consistency equations Eq.~\ref{eq:selfconsistent} for the pairing potential in real-space by
diagonalizing the BdG Hamiltonian Eq.~\ref{eq:BdG} implemented  on a
finite lattice with periodic boundary conditions~\cite{Zhu2016}. For the calculations
of the impurity bound state spectrum we choose a $31\times 31$ lattice
with the impurity located at the center. 

\begin{figure}[t]
\centering
\includegraphics[width=0.8\columnwidth, keepaspectratio]{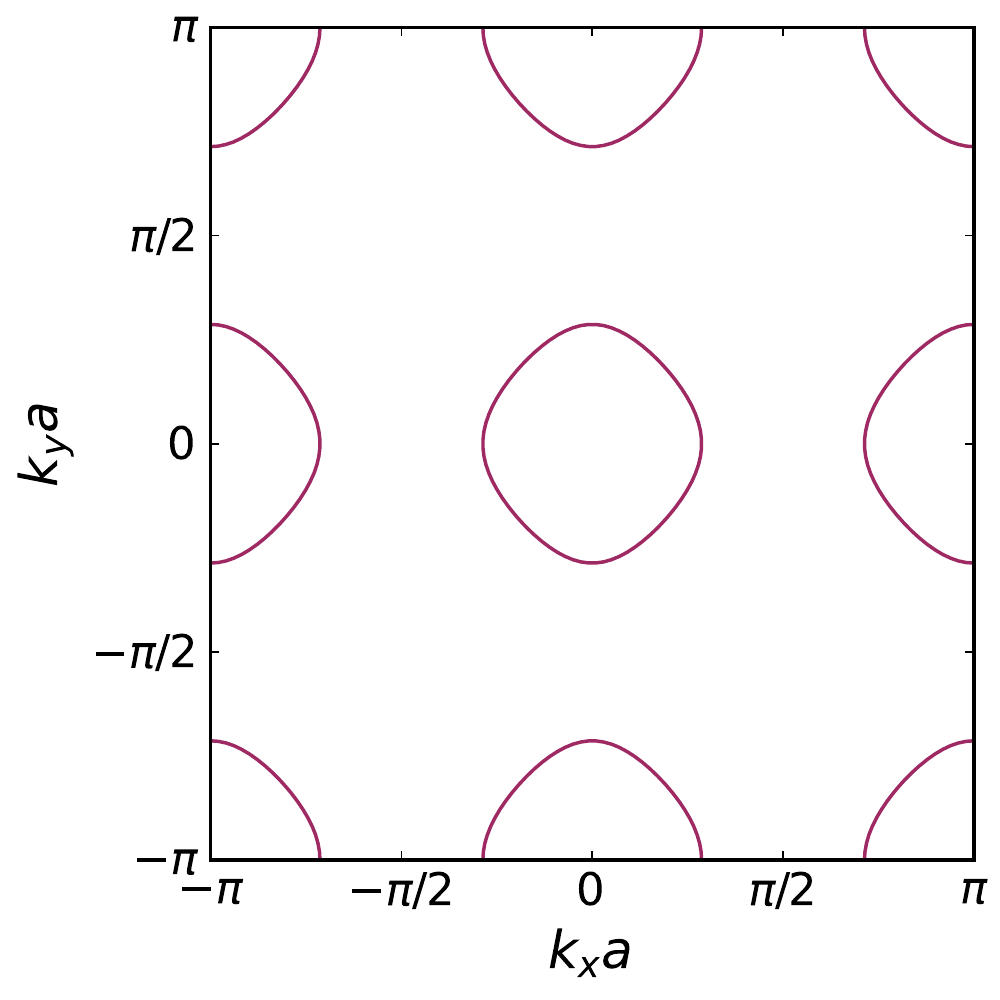}
\caption{\label{fig:fermi} The Fermi surface of the bilayer model with $k_{\text{F}} a =0.9$.}
\end{figure}

The four parameters $(t,\mu,t_\perp,\alpha)$ of our tight-binding
model describe a large variety of different band structures. Although
our primary aim is to understand the influence of the intertwined spin
and layer degrees of freedom on the impurity bound state spectrum, the
bound states can also be influenced by the shape of the Fermi surface
and the Fermi velocity~\cite{Ortuzar2022, Uldemolins2022}. To control for this variation we
observe that in the absence of intralayer hopping $t=0$, the iso-energy lines of the bands Eq.~\ref{eq:eigvals} in the Brillouin zone are the same for all parameter choices so long as $\alpha\neq 0$. Thus, for fixed filling, the shape of the Fermi surface is the same for all values of $t_\perp$ and $\alpha$. This is equivalent to setting the chemical potential to be
\begin{equation}
    \mu = \alpha \sqrt{ \text{sin}^2 ({k_{\text{F}} a})  + \left( \frac{t_\perp}{\alpha} \right)^2},
\end{equation}
for some fixed $k_{\text{F}}$; in the following we choose $k_{\text{F}}a = 0.9$ which
gives the Fermi surface shown in Fig. \ref{fig:fermi}. Since the
intralayer hopping appears with the identity matrix in
Eq.~\ref{eq:minbilayer}, it does not influence the gap at the Fermi
surface, and we thus do not expect that setting $t=0$ will qualitatively affect our results. This assumption will later be verified using the analytic $T$-matrix theory in Sec.~\ref{subsec:analytic}. 

The Fermi velocity in our model is given by
\begin{equation}
    {\mathbf v_k} = \frac{\alpha^2 a}{2\mu}\left(\sin(2k_xa)\hat{\bf x}+ \sin(2k_ya)\hat{\bf y}\right) \,,
\end{equation}
where $k_x$ and $k_y$ lie on the Fermi surface $\sin^2(k_xa) + \sin^2(k_ya)=\sin^2(k_{\text{F}}a)$. The magnitude of the Fermi velocity decreases with increasing $t_\perp$, which should reduce the coherence length even for fixed pairing potential, and may affect the impurity bound state spectrum. We keep the Fermi velocity constant 
by scaling $\alpha$ as
\begin{equation}
    \alpha = \frac{\alpha_0}{\sqrt{2}} \left( 1 + \sqrt{1 + \frac{4 t_\perp^2}{\alpha_0^2 \text{sin}^2 ({k_{\text{F}} a})}} \right)^{1/2},
\end{equation}
where $\alpha_0$ is the value of $\alpha$ at $t_\perp=0$. In the
following we set $\alpha_0$ as the unit of energy. For fixed
$k_{\text{F}}a$, the normal-state properties of the bilayer model now only
depend on the ratio $t_\perp/\alpha_0$. 

\begin{table*}[t]
\centering
\def\arraystretch{1.2}
 \begin{tabular}{|c|c|c|c|c|}  \hline
 description & matrix & irrep & effective gap (2D)& effective gap (3D) \\ \hline
 uniform intralayer singlet & $\Delta_0 \eta_0  i\sigma_y$ & $A_{\text{1g}}$ &
                                                                      $\Delta_0$
                                                  & $\Delta_0$ \\ \hline
 interlayer singlet & $\Delta_0\eta_x  i\sigma_y$ & $A_{\text{1g}}$ &
                                                              $\Delta_0 |\tilde{t}_\perp|$ & $\Delta_0 |\tilde{m}|$ \\ \hline
 staggered intralayer singlet & $\Delta_0\eta_z  i\sigma_y$ &
                                                             $A_{\text{2u}}$ &  $\Delta_0 \sqrt{1 - \tilde{t}_\perp^2}$ & $\Delta_0 \tilde{v}\sqrt{k_x^2+k_y^2}$\\ \hline
 interlayer triplet  & $\Delta_0\eta_y  \sigma_z i\sigma_y$ &
                                                             $A_{\text{1u}}$
                              &  $\Delta_0 \sqrt{1 -
                                \tilde{t}_\perp^2}$ & $\Delta_0
                                                      \sqrt{1 -
                                                      \tilde{m}^2}$ \\ \hline
  interlayer triplet  & $\{\Delta_0\eta_y  \sigma_x i\sigma_y,$ $\Delta_0\eta_y
                         \sigma_y i\sigma_y\}$ & $E_{\text{u}}$ &
                                                                  $\{\Delta_0
                                                                  \tilde{\alpha}|\sin(k_ya)|,
                                                                  \Delta_0
                                                                  \tilde{\alpha}|\sin(k_xa)|\}$
   & $\{\Delta_0
                                                                  \tilde{v}\sqrt{k_y^2+k_z^2},
                                                                  \Delta_0
                                                                  \tilde{v}\sqrt{k_x^2+k_z^2}\}$\\ \hline
\end{tabular}
\caption{\label{tab:pairing_states}Summary of possible $s$-wave
  pairing states in a bilayer superconductor. The columns give the
  description, the matrix form, the irreducible representation, and
  the value of the effective gap at the Fermi surface of the two models considered here. We adopt the short-hand notation $\tilde{t}_\perp = t_\perp/\mu$ and $\tilde{\alpha} = \alpha/\mu$.}
\end{table*}

The effective gap of the six $s$-wave pairing states at the Fermi
energy is given in Table.~\ref{tab:pairing_states}. All but the uniform
singlet depend on the parameters of the model, reflecting the
pair-breaking by the spin-layer texture of the normal-state band
wavefunctions. Importantly, we see that the $A_{\text{1g}}$, $A_{\text{1u}}$, and
$A_{\text{2u}}$ produce full isotropic gaps at the Fermi energy. In contrast,
the two $E_{\text{u}}$ states have point nodes. Since impurity bound states
are only well-defined in the presence of a full gap, we will not
examine the $E_{\text{u}}$ states within the mean-field theory; the impurity
virtual bound states which appear in these cases  will be analyzed using
the $T$-matrix theory developed below.  
Moreover, we will also not present results for the nontrivial $A_{\text{1g}}$ state, as this is essentially equivalent to the trivial $A_{\text{1g}}$ state when only a single band crosses the Fermi energy~\cite{Cavanagh2021}.
In solving the mean-field equations, the effective coupling constant $g$ in Eq.~\ref{eq:H_int} is chosen such that the pairing amplitude $\Delta_0 = \frac{g}{2} \langle \mathbf{c_{r_j}}\hat{\Delta}_\nu\mathbf{c_{r_j}}\rangle =0.08 \alpha_{0}$ in the absence of the impurity. 

\subsection{\label{subsec:analytic} $T$-matrix theory}

To complement the numerical mean-field calculation we also study the appearance of bound states at the impurity site within the $T$-matrix approximation. This has the benefit that it can be performed analytically if we relax the self-consistency requirement Eq.~\ref{eq:selfconsistent}, i.e. we treat $\Delta_0= \frac{g}{2} \langle \mathbf{c_{r_j}}\hat{\Delta}_\nu\mathbf{c_{r_j}}\rangle$ as independent of $\mathbf{r_j}$. 

The non-self-consistent $T$ matrix is defined as
\begin{equation}\label{eq:Tmatrix}
    T(i\omega_n) = \left[\hat{\mathbf{1}}-\hat{V}_{\text{imp}}G_0(i\omega_n) \right]^{-1} \hat{V}_{\text{imp}} \,,
\end{equation}
where $\hat{V}_{\text{imp}}$ is the impurity potential in Nambu
grading and 
$$
G_{0}(i\omega_n)=\frac{1}{N}\sum_{\bf k} G_{0}({\bf
  k},i\omega_n) \,,
$$ 
is the momentum-average of the Green's function $G_{0}({\bf
  k},i\omega_n)$ of the
superconductor in the absence of the impurity, with $N$ the number of
lattice points.

Analytic expressions for the Green's functions are complicated and
the exact evaluation of the momentum sum in Eq.~\ref{eq:Tmatrix} is difficult. To make progress, we replace the sum with an integral over energy near the Fermi surface and a Fermi-surface average
\begin{equation}
  \begin{aligned} \label{eq:sumtoint}
   \frac{1}{N}\sum_{\bf k} \to \int_{-\Lambda}^{\Lambda} N\left(\xi \right) d\xi \langle \ldots \rangle_{\text{FS}}.
  \end{aligned}
\end{equation}
where ${N}(\xi)$ is the density of states (DOS) and $\Lambda$ is an energy cutoff. Although analytic expressions for the DOS of the tight-binding model exist, here we pursue a more general approach by adopting a linear approximation to the DOS 
\begin{equation}
{N}\left(\xi \right)={N}_0+\xi {N}_0^\prime.
\end{equation}
where ${N}_0$ is the DOS at the Fermi energy and ${N}_0^\prime$ is
the first-order derivative with respect to energy. The parameter
${N}_0^\prime$ reflects the particle-hole asymmetry of the
DOS. Although the presence of particle-hole asymmetry is not essential for the formation of bound states, it is necessary for achieving
quantitative agreement with the tight-binding model. 

Using these approximations, the momentum-averaged Green's functions for the three pairing states are
\begin{widetext}
\begin{equation}\label{eq:Greensfunction}
G_0(i\omega_n) = -i\omega_n g_0(\eta_0+\tilde{t}_\perp \eta_x)\sigma_0\tau_0 - y(\eta_0+\tilde{t}_\perp \eta_x)\sigma_0\tau_z - \begin{cases} g_0\Delta_{\text{eff}}^{A_{\text{1g}}}(\eta_0 - \tilde{t}_\perp\eta_x) i\sigma_y \tau_x & A_{\text{1g}}\\
g_0\Delta_{\text{eff}}^{A_{\text{1u}}}\sqrt{1-\tilde{t}_\perp^2}\eta_y\sigma_x\tau_x & A_{\text{1u}}\\
g_0\Delta_{\text{eff}}^{A_{\text{2u}}}\sqrt{1-\tilde{t}_\perp^2}\eta_z\sigma_y\tau_y & A_{\text{2u}}
\end{cases} \,,
    \end{equation}
\end{widetext}
where $\tau_j$ are the Pauli matrices in Nambu space and its product
with $\eta$ and $\sigma$ matrices implies a Kronecker product, $\tilde{t}_\perp
= t_\perp/\mu$,  and 
\begin{equation}
    g_0 = \frac{{N}_0\pi}{2\sqrt{\omega_n^2+\Delta_{\text{eff}}^2}}\,.
\end{equation}
The parameter $y \approx
{N}^\prime_0 \Lambda$ encodes the particle-hole asymmetry, and we treat it as a fitting parameter. A sketch of the derivation of Eq.~\ref{eq:Greensfunction} is provided in appendix~\ref{sec:Appendix Green's function}. In particular, we show that Eq.~\ref{eq:Greensfunction} is valid for any band structure where only a single band crosses the Fermi energy. This gives us confidence that our
results hold beyond the restricted parameter space chosen for the
lattice model where we set $t=0$.

The relatively simple form of the integrated Green's function allows
us to evaluate the $T$ matrix Eq.~\ref{eq:Tmatrix} analytically. The impurity bound states are then determined by making the analytic continuation $i\omega_n \rightarrow \omega + i0^{+}$ and solving for the poles at subgap energies $|\omega|<\Delta_{\text{eff}}$. 
In the following, we express our results in turns of dimensionless quantities $\tilde{\omega}=\omega/|\Delta_{\text{eff}}|$, $\tilde{V}=\frac{\pi}{2}N_{0}V$ and $\tilde{y}=y/(\frac{\pi}{2}N_{0})$.

\begin{table*}[t]
\centering
\def\arraystretch{1.2}
    \begin{tabular}{|c|c|c|c|c|c|c|c|c|c|c|c|c|c|c|c|c|c|}  \hline
    & & \multicolumn{16}{c|}{impurity} \\ \hline
    & & $\eta_0\sigma_0$ & $\eta_x\sigma_0$ & $\eta_y\sigma_0$ & $\eta_z\sigma_x$ & $\eta_z\sigma_y$ & $\eta_z\sigma_z$ & $\eta_z\sigma_0$ & $\eta_y\sigma_x$ & $\eta_y\sigma_y$ & $\eta_y\sigma_z$ & $\eta_x\sigma_x$ & $\eta_x\sigma_y$ & $\eta_x\sigma_z$ & $\eta_0\sigma_x$ & $\eta_0\sigma_y$ & $\eta_0\sigma_z$ \\ \hline\hline
    \multirow{2}{*}{symmetry} & $\mathcal{T}$ & \tickYes & \tickYes & \tickNo & \tickNo & \tickNo & \tickNo & \tickYes & \tickYes & \tickYes & \tickYes & \tickNo & \tickNo & \tickNo & \tickNo & \tickNo & \tickNo \\ \cline{2-18}
    & $\mathcal{PT}$ & \tickYes & \tickYes & \tickYes & \tickYes & \tickYes & \tickYes & \tickNo & \tickNo & \tickNo & \tickNo & \tickNo & \tickNo & \tickNo & \tickNo & \tickNo & \tickNo \\ \hline\hline
    %& $\mathcal{P}$ & \tickYes & \tickYes & \tickNo & \tickNo & \tickNo & \tickNo & \tickNo & \tickNo & \tickNo & \tickNo & \tickYes & \tickYes & \tickYes & \tickYes & \tickYes & \tickYes \\ \hline
    \multirow{5}{*}{fitness} & $A_{\text{1g}}$ & $+1$ & $+1$ & $-1$ & $-1$ & $-1$ & $-1$ & $+1$ & $+1$ & $+1$ & $+1$ & $-1$ & $-1$ & $-1$ & $-1$ & $-1$ & $-1$ \\ \cline{2-18}
    & $A_{\text{1u}}$ & $+1$ & $-1$ & $-1$ & $-1$ & $-1$ & $+1$ & $-1$ & $-1$ & $-1$ & $+1$ & $-1$ & $-1$ & $+1$ & $+1$ & $+1$ & $-1$ \\ \cline{2-18}
    & $A_{\text{2u}}$ & $+1$ & $-1$ & $+1$ & $-1$ &
                                                                  $-1$ & $-1$ & $+1$ & $-1$ & $-1$ & $-1$ & $+1$ & $+1$ & $+1$ & $-1$ & $-1$ & $-1$ \\\cline{2-18}
    & $E_{\text{u},x}$ & $+1$ & $-1$ & $-1$ & $-1$ &
                                                                  $+1$ & $-1$ & $-1$ & $-1$ & $+1$ & $-1$ & $-1$ & $+1$ & $-1$ & $+1$ & $-1$ & $+1$ \\\cline{2-18}
    & $E_{\text{u},y}$ & $+1$ & $-1$ & $-1$ & $+1$ &
                                                                  $-1$
                                                               &
                                                                 $-1$ & $-1$ & $+1$ & $-1$ & $-1$ & $+1$ & $-1$ & $-1$ & $-1$ & $+1$ & $+1$ \\ \hline\hline

    \multirow{5}{*}{classification} & $A_{\text{1g}}$ & - & - & a* & a* & a* & a* & - & - & - & - & b* & b* & b* & b* & b* & b* \\ \cline{2-18}
    & $A_{\text{1u}}$ & a & b & d & d & d & c & d & d & d & c & f & f & e & e & e & f \\ \cline{2-18}
    & $A_{\text{2u}}$ & a & b & c & d & d & d & c & d & d & d & e & e
                                                                                                                                                                                                    &
                                                                                                                                                                                                      e & f & f & f \\ \cline{2-18}
& $E_{\text{u},x}$ & a & b & d & d & c & d & d & d & c & d & f & e & f & e & f & e \\
 \cline{2-18}
& $E_{\text{u},y}$ & a & b & d & c & d & d & d & c & d & d & e & f & f
                                                                                                                                                                                                                       & f & e & e \\ \hline
    \end{tabular}
\caption{Categorization of bound state spectra for the $s$-wave
  states considered in this paper. 
The symmetry row shows whether the impurity preserves/breaks
 (\tickYes\color{black}/\tickNo\color{black})
time-reversal symmetry $\mathcal{T}$ and combined
parity-time-reversal symmetry $\mathcal{PT}$. The fitness row
indicates the solution for $\lambda$ in Eq.~\ref{eq:genAnderson}, with
$\lambda=+1$ ($-1$) indicating a fit (unfit) impurity for each pairing
state. 
The final row shows the classification of the bound states. For the
$A_{\text{1g}}$ pairing, bound
states only appear for time-reversal symmetry-breaking impurities, and
we find two categories a${}^\ast$ and b$^{\ast}$, respectively corresponding to the
panels (a) and (b) in Fig.s~\ref{fig:singlet_Vimp}
and~\ref{fig:singlet_tperp}. For the odd-parity $A_{\text{1u}}$ and $A_{\text{2u}}$
states we find six different categories a-f, corresponding to the panels
(a)-(f) in Fig.s~\ref{fig:evals_vs_Vimp}
and~\ref{fig:evals_vs_t_perp}. For the $A_{\text{2u}}$ and $E_{\text{u}}$ pairing
states, this classification also holds for the virtual bound states in
Fig.~\ref{fig:LDOS}.}\label{tab:imp_survey}
\end{table*}

\section{\label{sec:boundstates}Bound states}

The central result of our work is that the bound states which form
around impurities fall into a number of distinct classes, which
exhibit qualitatively different behavior as a function of interlayer
coupling $\tilde{t}_\perp$ and impurity strength $\tilde{V}$. The
classification of the 16 possible 
momentum-independent impurities for the different pairing states
is summarized in Table \ref{tab:imp_survey}, and is based on whether
the impurity preserves time-reversal ($\mathcal{T}$) or combined
parity-time-reversal ($\mathcal{PT}$) symmetries, and the fitness of
the impurity with respect to the pairing state. 
$A_{\text{1g}}$ interlayer pairing this gives rise to three distinct
categories of the spectrum, whereas for the $A_{\text{1u}}$ and $A_{\text{2u}}$
states there are six categories.  

Before discussing the distinct cases in detail, we first comment on the significance of time-reversal ($\mathcal{T}$) or combined parity-time-reversal ($\mathcal{PT}$) symmetry. As these are both antiunitary symmetries, Kramers' theorem dictates that the spectrum is two-fold degenerate when either is present. Since the clean BdG Hamiltonian is symmetric under both time-reversal and inversion, their presence or absence is determined entirely by the impurity.

\subsection{\label{sub:intralayer_singlet} $A_{\text{1g}}$ intralayer singlet pairing}

\begin{figure}[]
\centering
\hspace{-0.16in}
  \begin{overpic}[width=1.03\columnwidth, keepaspectratio]{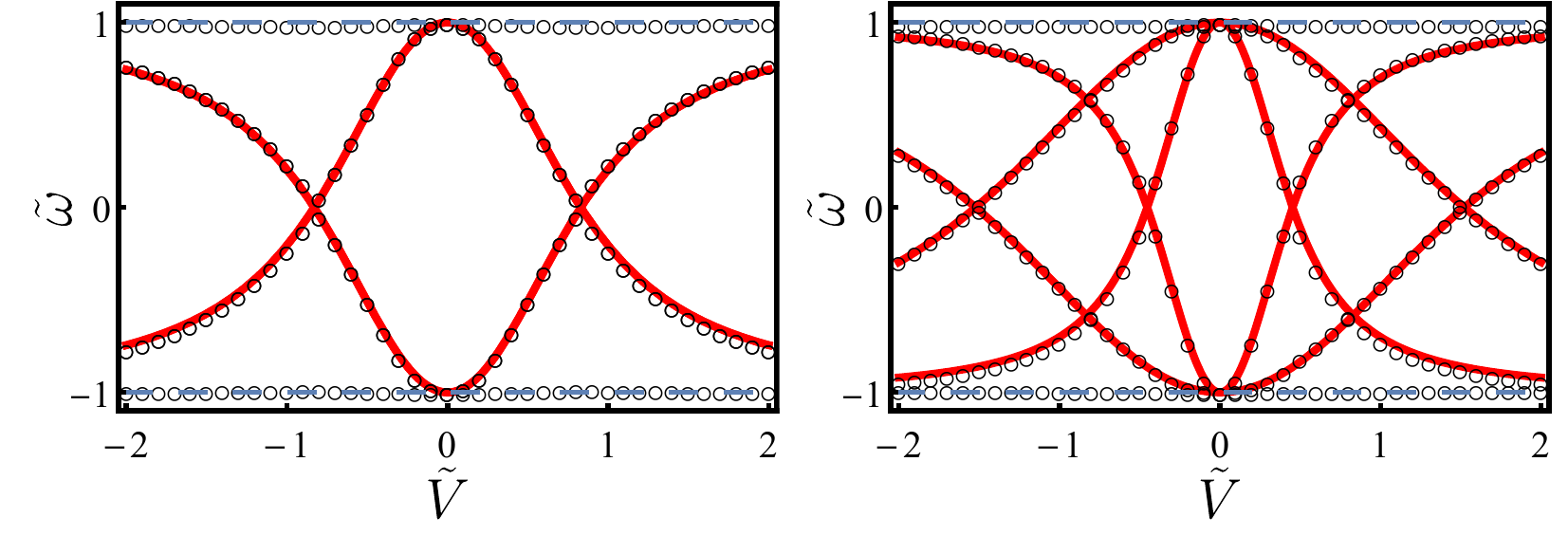}
    \put(3, 2){(a)}
    \put(50, 2){(b)}
  \end{overpic}
  \vspace{-0.2in}
\caption{\label{fig:singlet_Vimp} Subgap spectrum as a function of
  the dimensionless impurity strength $\tilde{V}$ for the 
  intralayer singlet $A_{\text{1g}}$ pairing. Plots (a) and (b) correspond to the
  impurity classifications for this pairing state
  tabulated in Table~\ref{tab:imp_survey}. The circles are the results
  of the self-consistent mean-field theory. The dashed grey line is
  the analytic prediction for the gap edge, and the solid red line
  represents the results of the analytic $T$-matrix theory for the
  subgap states. The interlayer impurity hopping is set as 
$\tilde{t}_\perp=0.54$ and the particle-hole asymmetric strength is set as $\tilde{y}=0.50$.}
\end{figure}

\begin{figure}[]
\centering
\hspace{-0.16in}
  \begin{overpic}[width=1.03\columnwidth, keepaspectratio]{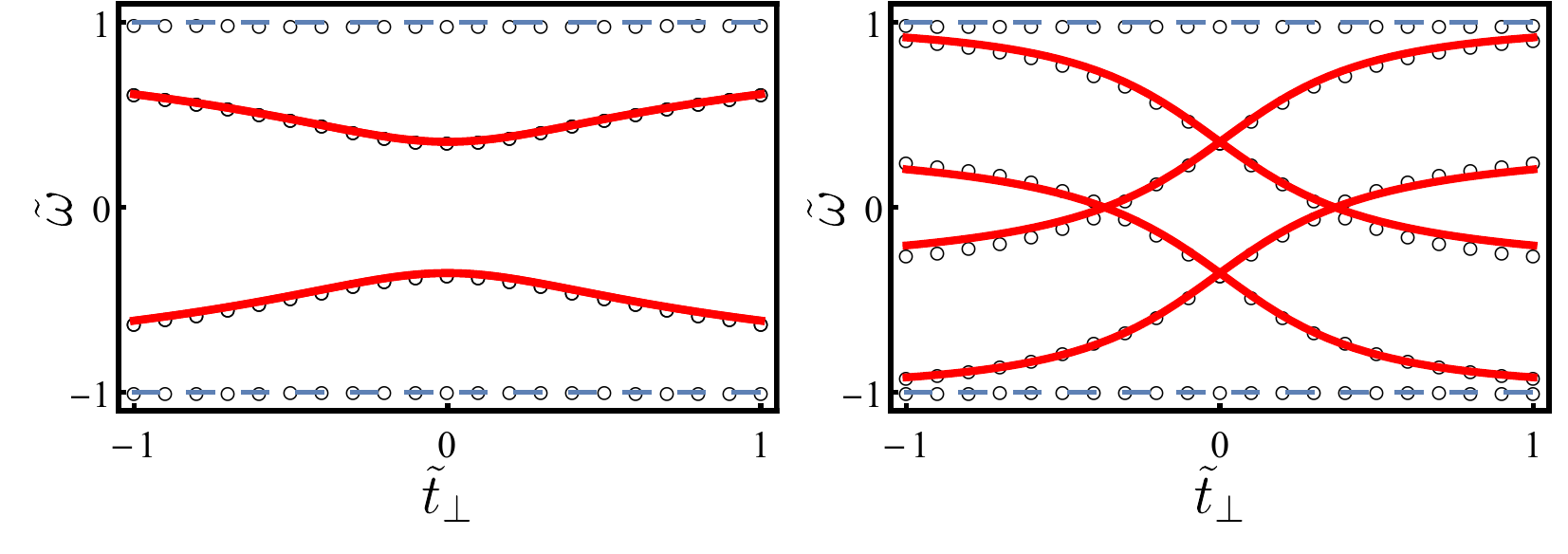}
    \put(3, 2){(a)}
    \put(50, 2){(b)}
  \end{overpic}
  \vspace{-0.2in}
\caption{\label{fig:singlet_tperp} Subgap spectrum as a function of
the dimensionless interlayer hopping $\tilde{t}_\perp$
for the intralayer singlet $A_{\text{1g}}$ pairing. Plots (a)-(d)
correspond to the four possible impurity classifications for this
pairing state tabulated in Table~\ref{tab:imp_survey}. The
normalized impurity strength is set as $\tilde{V}=0.64$ and the particle-hole asymmetric strength is set as $\tilde{y}=0.50$. The same style is used as in
Fig. \ref{fig:singlet_Vimp}.}
\end{figure}

The intralayer singlet $A_{\text{1g}}$ state pairs electrons in time-reversed
states, and is thus subject to Anderson's theorem. Since the normal
state Hamiltonian has time-reversal symmetry, this state is perfectly
fit in the clean limit and there is only intraband pairing. The only
pair-breaking impurities are those that break time-reversal symmetry,
and we accordingly only find bound states around the impurities in
these cases, as shown as a function of impurity strength and
interlayer coupling in
Fig.~\ref{fig:singlet_Vimp} and \ref{fig:singlet_tperp},
respectively. These figures also show the excellent quantitative
agreement between the mean-field and the $T$-matrix theories. Analytic
expressions for the bound
states obtained via the $T$-matrix method are complicated and are
given in the appendix~\ref{sec:Appendix A1g BS}. 

Our results reveal two distinct classes of bound states: class (a)
impurities where the bound states are two-fold degenerate, and class
(b) impurities where the bound states are not degenerate. The
degeneracy of the states in case (a) is ensured by a generalized
Kramers' theorem based on the preserved $\mathcal{PT}$ symmetry; in
contrast, in case (b) both $\mathcal{T}$ and $\mathcal{PT}$ symmetries
are broken, and there is no symmetry enforcing degeneracy. We note
that even in case (b) the spectra is two-fold 
degenerate in the case of vanishing interlayer coupling.

\subsection{\label{sub:interlayer_triplet} Odd-parity $A_{\text{1u}}$ and $A_{\text{2u}}$ states}

We now turn to the bound state spectrum in the case of the fully-gapped odd-parity states, which is considerably richer than the $A_{\text{1g}}$ state. The classification of the spectra is the same for the two odd-parity states, reflecting the fact that in the clean limit the BdG Hamiltonian is the same up to unitary transformation, specifically we have
\begin{equation}
{\cal H}_{\text{BdG},A_{\text{2u}}} = U^\dagger{\cal H}_{\text{BdG},A_{\text{1u}}}U \label{eq:A1utoA2umap} \,,
\end{equation}
where ${\cal H}_{\text{BdG},A_{\text{1u}}(A_{\text{2u}})}$ is the 
BdG Hamiltonian matrix for the $A_{\text{1u}}$ ($A_{\text{2u}}$)
state in the absence of the impurity, and
\begin{equation}
U = \begin{pmatrix}\exp(i\frac{\pi}{4}\eta_x\sigma_z) & 0 \\0 &
  \exp(-i\frac{\pi}{4}\eta_x\sigma_z)\end{pmatrix}. \label{eq:UA1utoA2u}
  \end{equation}
The impurity potential is
generally not invariant under this unitary transform: this allows us
to map each impurity potential in the $A_{\text{1u}}$ pairing state to
another impurity potential in the $A_{\text{2u}}$ pairing state. It also
follows from Eq.~\ref{eq:A1utoA2umap} that the fitness of the $A_{\text{1u}}$
and $A_{\text{2u}}$ states are the same in the clean limit, with both pairing
states unfit with respect to the interlayer coupling $t_\perp$. Both
states are perfectly fit when $t_\perp=0$, and in this limit there also
exists a unitary transformation that maps the odd-parity state to the
$A_{\text{1g}}$ intralayer singlet state~\cite{FuBerg2010}. We therefore
expect that the bound states for the odd-parity $s$-wave pairing will strongly depend on the interlayer coupling.

\begin{figure}[]
\centering
\hspace{-0.16in}
  \begin{overpic}[width=1.03\columnwidth, keepaspectratio]{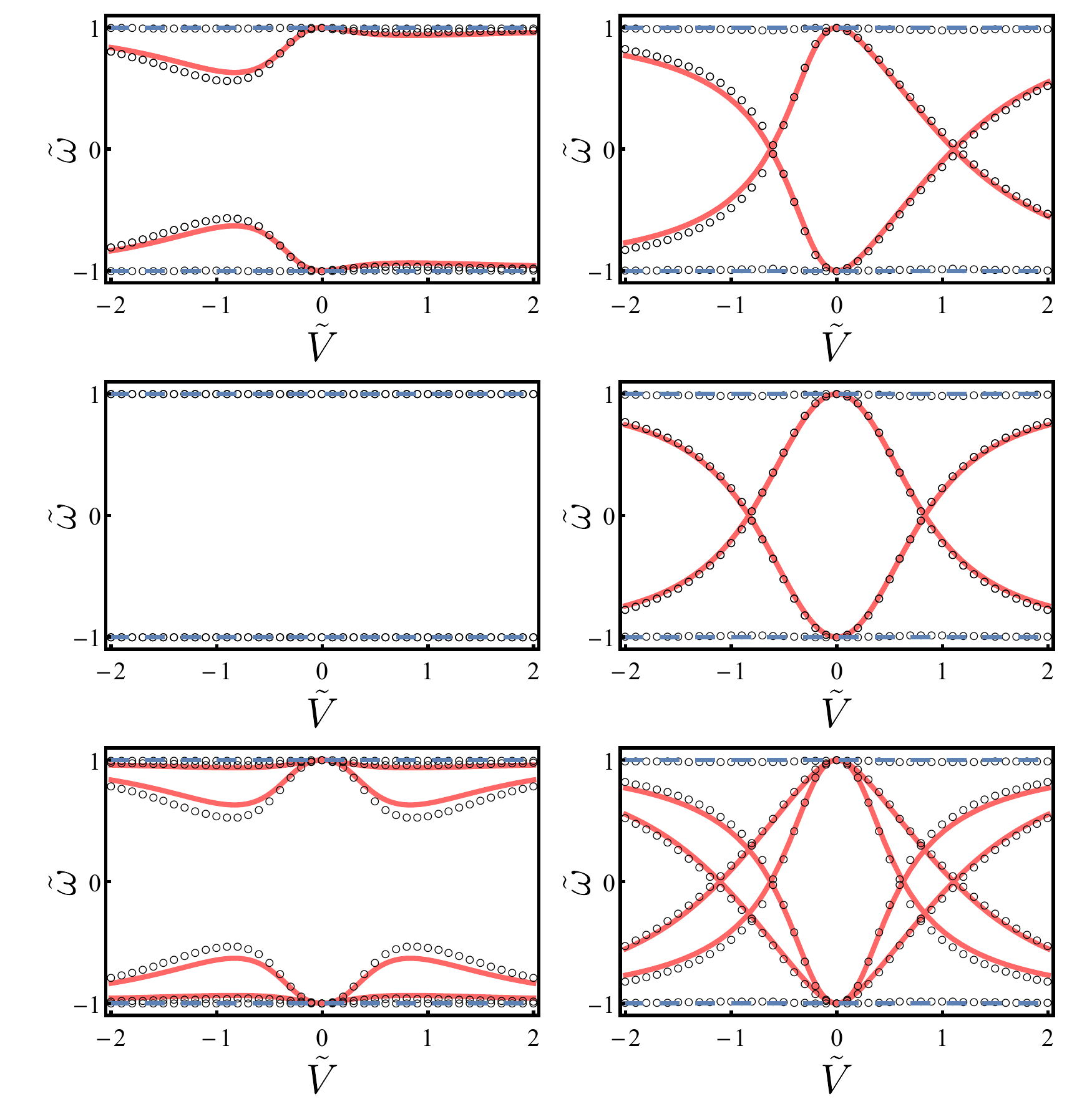}
    \put(3,  67){(a)}
    \put(50, 67){(b)}
    \put(3, 34.5){(c)}
    \put(50, 34.5){(d)}
    \put(3, 2){(e)}
    \put(50, 2){(f)}
  \end{overpic}
  \vspace{-0.15in}
\caption{\label{fig:evals_vs_Vimp} Subgap spectrum as a function of
  the dimensionless impurity strength $\tilde{V}$ for the odd-parity $A_{\text{1u}}$ or $A_{\text{2u}}$ pairing states. Plots (a)-(f) correspond to the six possible impurity classes summarized in Table \ref{tab:imp_survey}. The interlayer impurity hopping is set as $\tilde{t}_\perp=0.54$ and the particle-hole asymmetric strength is set as $\tilde{y}=0.50$. The same style is used as in Fig. \ref{fig:singlet_Vimp}.}
\end{figure}

\begin{figure}[]
\centering
\hspace{-0.16in}
  \begin{overpic}[width=1.03\columnwidth, keepaspectratio]{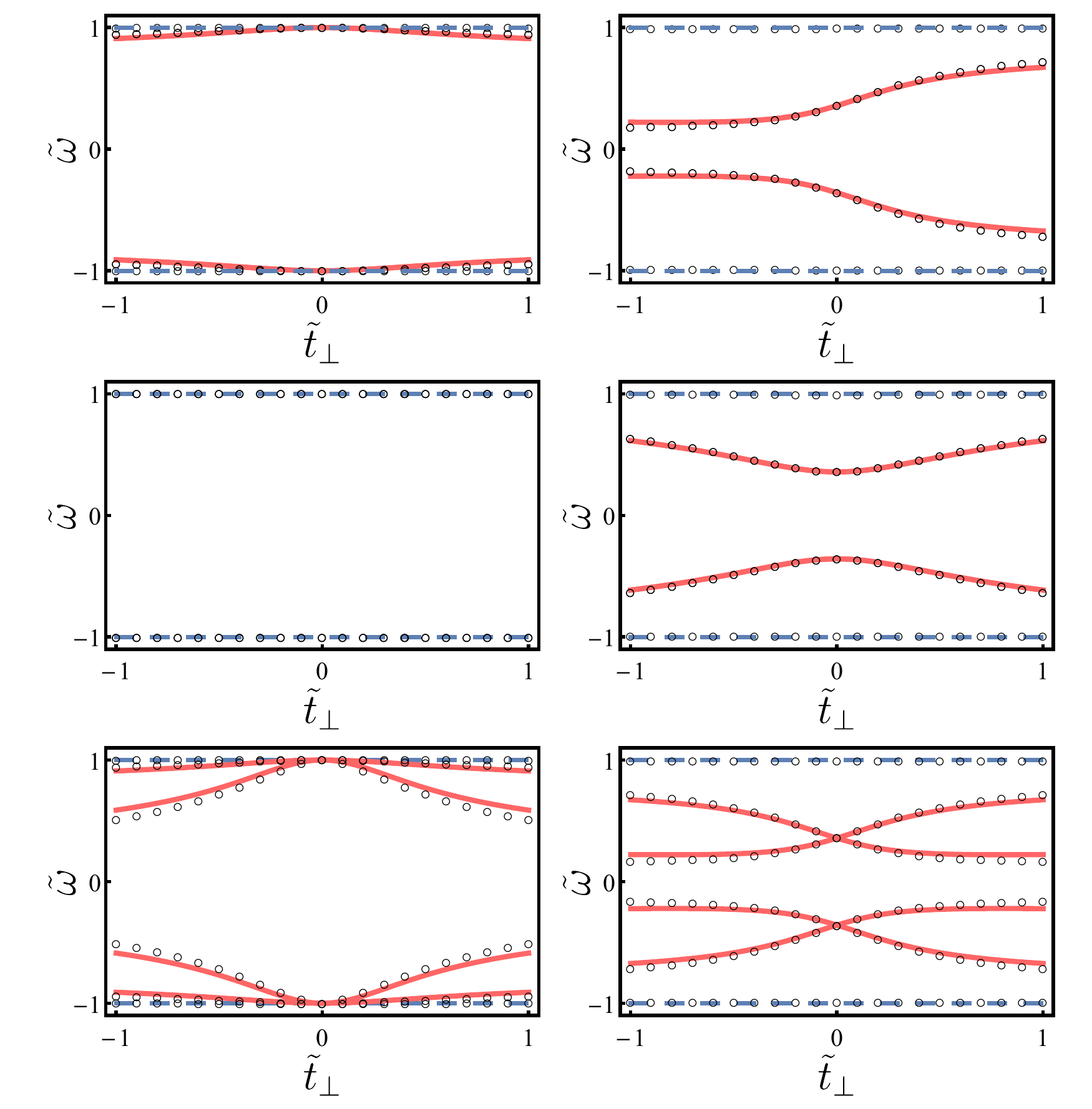}
    \put(3,  67){(a)}
    \put(50, 67){(b)}
    \put(3, 34.5){(c)}
    \put(50, 34.5){(d)}
    \put(3, 2){(e)}
    \put(50, 2){(f)}
  \end{overpic}
  \vspace{-0.15in}
\caption{\label{fig:evals_vs_t_perp} Subgap spectrum as a function of the dimensionless interlayer hopping $\tilde{t}_\perp$ for the odd-parity $A_{\text{1u}}$ or $A_{\text{2u}}$ pairing states. Plots (a)-(f) correspond to the six possible impurity classes summarized in Table \ref{tab:imp_survey}. The normalized impurity strength is set as $\tilde{V}=0.64$ and the particle-hole asymmetric strength is set as $\tilde{y}=0.50$. The same style is used as in Fig. \ref{fig:singlet_Vimp}.}
\end{figure}
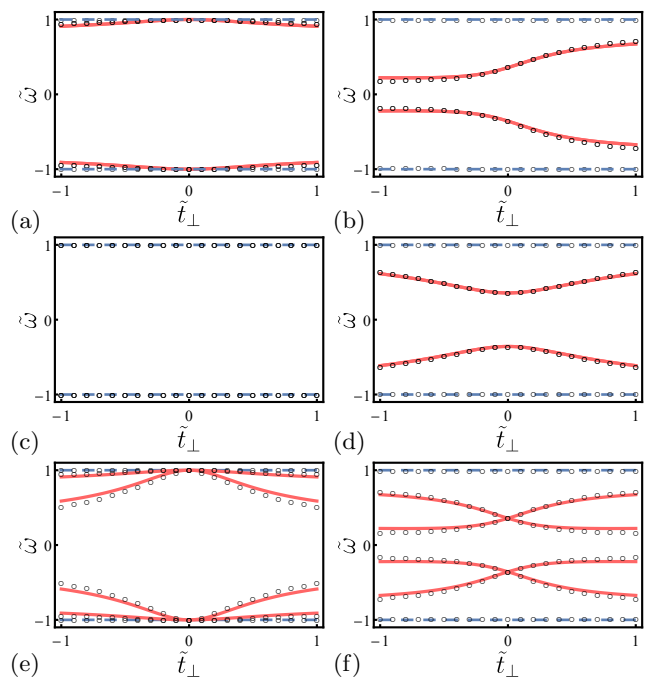

We find six different classes of bound state spectra, which are shown in Figs.~\ref{fig:evals_vs_Vimp} and \ref{fig:evals_vs_t_perp}
as a function of $\Tilde{V}$ and $\Tilde{t}_\perp$, respectively. There is
again excellent agreement between the self-consistent real-space
lattice theory and the continuum $T$-matrix approximation, using
the same  particle-hole asymmetry parameter $\tilde{y}=0.50$ as for the $A_{\text{1g}}$ results. The dependence on the impurity strength shows a clear difference between the fit impurities (cases (a), (c), and (e))
and the unfit impurities (cases (b), (d), and (f)): although bound
states are generally present, only for the unfit impurities
do the bound states cross zero energy as a function of impurity strength. The closing of the gap by the impurity evidence the pair-breaking nature of the unfit impurities. The role of the fitness with respect to the normal-state Hamiltonian is clearly revealed by the dependence on the interlayer coupling $t_\perp$ shown in Fig.~\ref{fig:evals_vs_t_perp}: bound states for the fit impurities  are always absent for vanishing intralayer coupling, as anticipated by the mapping between the odd-parity and the $A_{\text{1g}}$ states.

Greater insight is provided by the analytic
expressions for the bound states found in the $T$-matrix theory. These
expressions are generally rather complicated, so to more clearly
reveal the relationship we set $\tilde{y}=0$; general
expressions with $\tilde{y}\neq0$ are provided in
Appendix~\ref{sec:Appendix six BS}.  
In the particle-hole-symmetric limit, the bound states are 
\begin{equation}\label{eq:analytic y=0}
\tilde{\omega} = \pm\frac{\lambda+
\left(1-\tilde{t}_\perp^2\right) \tilde{V}^2}{
  \sqrt{ 1 + 2 (1 + S_{t_\perp} \tilde{t}_\perp^2) \tilde{V}^2 + ( 1-\tilde{t}_\perp^2)^2 \tilde{V}^4}}\,,
\end{equation}
where $\lambda=\pm 1$ is the solution to Eq.~\ref{eq:genAnderson} (i.e. the fitness of the impurity), and the parameter $S_{t_\perp}=+1$ ($-1$) according as the impurity commutes (anticommutes) with the interlayer hopping term.
The combinations of $(S_{t_\perp},\lambda)$ give rise to
four distinct bound state spectra: $(1,1)$ corresponding to cases (a)
and (e), $(1,-1)$ corresponding to (b) and (f), $(-1,1)$ corresponding
to (c), and $(-1,-1)$ corresponding to (d). The difference between the
(a) and (e) (or (b) and (f)) cases is only apparent for nonzero
asymmetry parameter $\tilde{y}$. In particular, a nonzero value of the
asymmetry parameter is necessary to lift the two-fold degeneracy of
the bound states, and is also responsible for the asymmetric
dependence on the impurity potential seen in cases (a) and (b), and
the interlayer coupling strength in case (b). These effects of the
asymmetry parameter are consistent with previous work~\cite{Wang2012, Mashkoori2017}.

Despite these limitations, Eq.~\ref{eq:analytic y=0} nevertheless
reveals the role of fitness. We first consider the different
impurity fitness cases, $\lambda=\pm 1$: in the case of
the fit impurity $\lambda=1$, the numerator of
Eq.~\ref{eq:analytic y=0} is always positive and so the impurity
cannot close the gap; on the other hand, for unfit impurities
$\lambda=-1$ the numerator vanishes at a critical value
$\tilde{V} = 1/\sqrt{1 - \tilde{t}_{\perp}^2}$, thus
closing the gap. We also observe that the bound-state energies depend
nontrivially on the interlayer coupling, reflecting the interplay of
the impurity with the spin and layer degrees of freedom of the clean
Hamiltonian, and thus the normal-state fitness of the pairing
potential. 
The dependence on $S_{t_\perp}$ is generally more subtle, but in the
case $S_{t_\perp}=-1$ it ensures that the term under the square root
is a complete square. For $\lambda=1$ this cancels the
numerator, and thus there are no subgap states due to the impurity. 

The parameters $S_{t_\perp}$ and $\lambda$ also enter the Born approximation result for the suppression of the critical temperature by the presence of a finite concentration $n_{\text{imp}}$ of impurities~\cite{Cavanagh2020,Cavanagh2021}. Here the critical temperature is given by the solution of the equation
\begin{equation}
  \log\left(\frac{T_c}{T_{c0}}\right) = \Psi\left(\frac{1}{2}\right) - \Psi\left(\frac{1}{2} + \frac{1}{4\pi k_B T_c \tau}\right) \,,
\end{equation}
where $T_{c0}$ is the critical temperature in the clean limit and the effective scattering rate is
\begin{equation}
  \tau^{-1} = \pi n_{\text{imp}} |V|^2 {N}_0 \left(1
    + S_{t_\perp}\tilde{t}_\perp^2 -
    \lambda\langle 1-\tilde{F}_{\nu,{\bm k}}\rangle_{\text{FS}}\right)\,. \label{eq:tau} 
\end{equation}
The first two terms inside the brackets give the normal-state
scattering rate, while the bulk fitness parameter Eq.~\ref{eq:fitnessnorm} controls the
effective scattering rate in the superconducting state.
For the unfit impurities with $\lambda=-1$ the effective
scattering rate is higher than the normal-state scattering rate, and the critical
temperature is thus suppressed by increasing impurity concentration
faster than the Abrikosov-Gor'kov predictions for an odd-parity state~\cite{Mineev1999};
Conversely, the fit impurities with $\lambda=1$
have a lower effective scattering rate than in the normal-state, and so the critical temperature displays a weaker suppression than predicted by the usual theory. 
Intriguingly, in case (c) with
$(S_{t_\perp},\lambda)=(-1,1)$, the scattering rate vanishes, implying that there is no suppression of the critical temperature. 
Our results are consistent with the observation that the presence of
impurity bound states, even when produced by the fit impurities, indicates some degree of pair-breaking. 

\subsection{ Sublattice localized impurities}

A general impurity potential may involve a linear combination of all the matrices listed in Tab.~\ref{tab:imp_survey}, and hence includes both fit and unfit components. 
%In the Born approximation, the total scattering rate is obtained by adding or subtracting the scattering rate due to each component according as it is fit or unfit~\cite{Cavanagh2021}. 
Since the formation of bound states is a nonlinear effect we cannot directly infer the bound state spectrum of a general impurity potential from the spectra produced by each component. A physically-interesting case of a multi-component impurity has the impurity potential localized on a single sublattice, i.e. arising from the replacement of an atom at a particular sublattice site by another species. The corresponding impurity potential has the matrix form $(\eta_{0} \pm \eta_{z})\sigma_{\mu}$, which involves two matrices which generally belong to different classes in Tab.~\ref{tab:imp_survey}. 

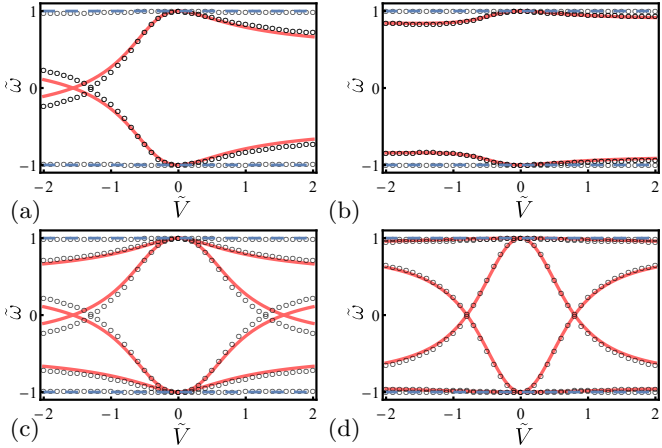
\begin{figure}[]
\centering
\hspace{-0.16in}
  \begin{overpic}[width=1.03\columnwidth, keepaspectratio]{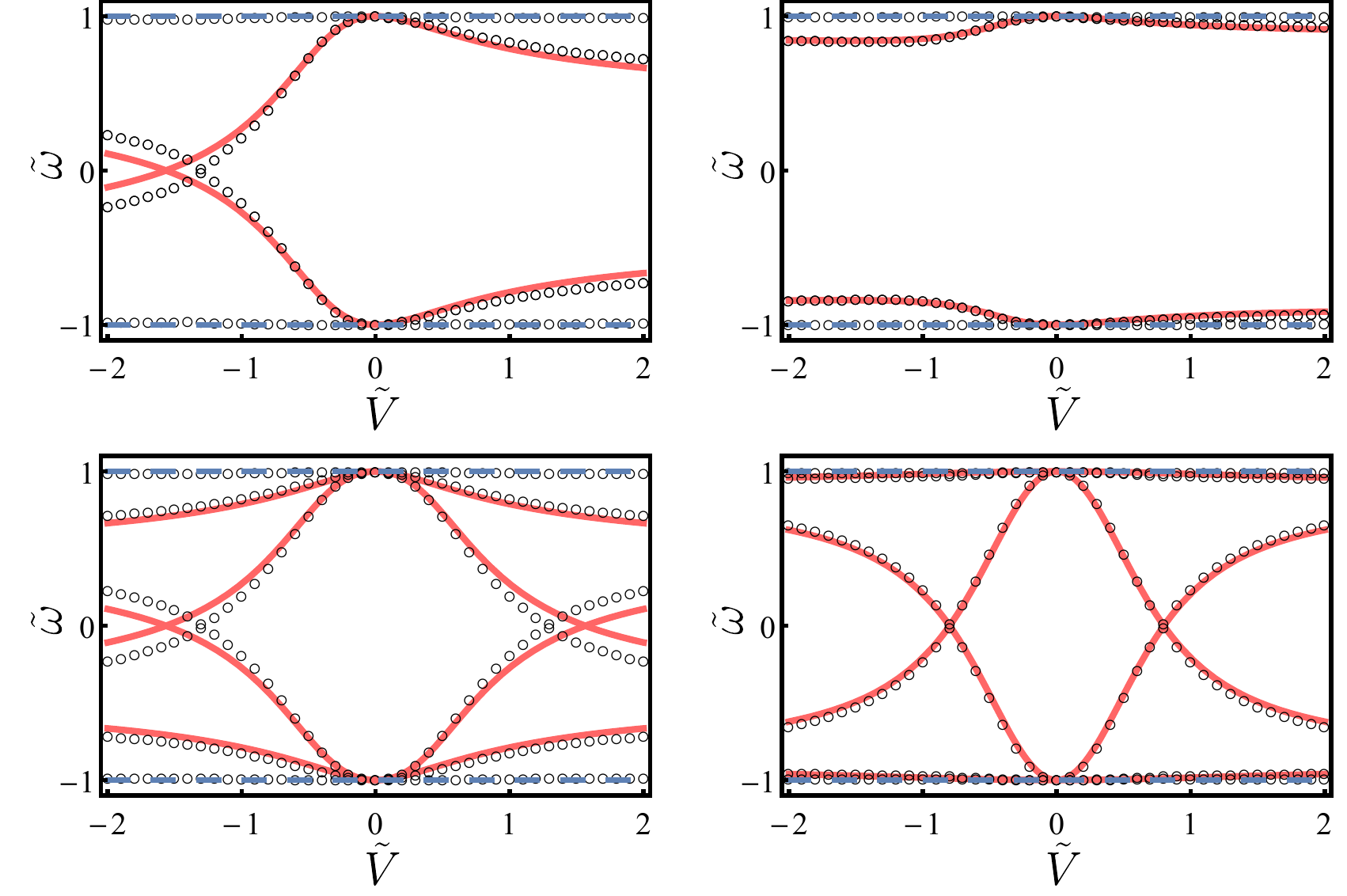}
    \put(3, 34.5){(a)}
    \put(50, 34.5){(b)}
    \put(3, 2){(c)}
    \put(50, 2){(d)}
  \end{overpic}
  \vspace{-0.15in}
\caption{\label{fig: sublattice impurities} Subgap spectrum for sublattice-localized impurities as a function of the dimensionless impurity strength $\tilde{V}$ for odd-parity $A_{\text{1u}}$ (left) or $A_{\text{2u}}$ (right) pairing states. The top (bottom) row corresponds to potential (magnetic) impurities localized to a single sublattice. The interlayer impurity hopping is set as $\tilde{t}_\perp=0.54$ and the particle-hole asymmetric strength is set as $\tilde{y}=0.50$. The same style is used as in Fig. \ref{fig:singlet_Vimp}.}
\end{figure}

In Fig.~\ref{fig: sublattice impurities} we plot the subgap spectrum for sublattice-localized potential ($\mu=0$) and magnetic ($\mu=x$, $y$, $z$) impurities in the two odd-parity states. For both pairing states the subgap spectrum is independent of the polarization of the magnetic impurity.
For the $A_{2u}$ state, the two components of the potential (magnetic) impurity are both fit (unfit). Consistent with our argument above, only the the subgap states of the unfit magnetic impurity cross zero energy.

The situation with the $A_{1u}$ state is more subtle, as for both potential and magnetic impurities one component is fit and the other is unfit. In the particle-hole-symmetric limit (i.e. $\tilde{y}=0$), the subgap states asymptotically go to zero at infinite impurity strength; switching on particle-hole asymmetry shifts the zero-crossing to a finite value of $\tilde{V}$. Thus, the presence of the unfit impurity potential is still consistent with the appearance of a zero-crossing of the subgap states.

\section{\label{sec:virtual bound states}Three-dimensional model}

The full gap of the $A_{\text{1u}}$ and $A_{\text{2u}}$ states in the
two-dimensional tight-binding model is convenient for
numerical calculations. However, in three dimensions the $A_{\text{2u}}$
state generically has point nodes. Due to hybridization with the
low-lying quasiparticle states in the 
vicinity of these nodes, the impurity bound states found above are
replaced by resonances or so-called virtual bound
states~\cite{Zhu2016,Balatsky1995,BalatskyRMP2006}. Although it is
numerically prohibitive to study the subgap spectra of the
three-dimensional system  
using the tight-binding mean-field theory, the non-self-consistent
$T$-matrix approximation can be readily applied to this problem.
Considering the excellent agreement between 
the mean-field theory and the $T$-matrix approximation observed above for
the two-dimensional model, we expect that the $T$-matrix approximation
will also give accurate results in three dimensions.

\subsection{Minimal model}

Close to the $\Gamma$ point, and keeping terms up to linear order in
momentum, a minimal three-dimensional generalization of our normal-state Hamiltonian is
\begin{align}
H({\bm k})  = & -\mu\eta_0\sigma_0 + m\eta_x\sigma_0 +
    v_zk_z\eta_y\sigma_0 \notag \\
& + v k_x\eta_z\sigma_y - vk_y\eta_z\sigma_x \,,\label{eq:Ham3D}
\end{align}
where $m$ is the mass term, $v$ is the in-plane velocity, and $v_z$ is
the out-of-plane velocity. The presence of the out-of-plane velocity
$v_z$ extends the model to three dimensions; scaling the $z$-component
of the momentum allows us to eliminate the velocity anisotropy,
i.e. in the following we set $v_z=v$. In terms of the parameters of
the tight-binding model we have $m = t_\perp$ and $v = \alpha
a$. We note that a term proportional to $\eta_z\sigma_z$ is
allowed by symmetry, but we neglect it as it appears with much
higher power of momenta: for the tetragonal system considered here it
requires taking the momentum expansion to the fifth order, whereas for 
trigonal or hexagonal systems it appears at the third order. 
 
\subsection{$A_{\text{1g}}$ and $A_{\text{1u}}$ pairing states}

The momentum-averaged Green's function for both the $A_{\text{1g}}$ and
$A_{\text{1u}}$ pairing states in the three-dimensional model is the same as
in Eq.~\ref{eq:Greensfunction}, and thus the
conclusions of Sec.~\ref{sec:boundstates} remain valid.
We note that including the $\eta_z\sigma_z$ term in the
Hamiltonian does introduce some gap anisotropy for the $A_{\text{1u}}$
state and may therefore modify our
results. So long as this term is small compared to the other terms in
Eq.~\ref{eq:Ham3D}, however, we expect that the effects on the bound
state spectrum will be small.

\subsection{$A_{\text{2u}}$ pairing state}

The gap opened by the $A_{\text{2u}}$ pairing state on a three-dimensional
Fermi surface is required by symmetry to have a node along the $k_z$
axis~\cite{Mineev1999}. In our model, the gap at the Fermi energy is given by
\begin{align}
  \Delta_{A_{\text{2u}}}({\bm k}) = &\Delta_{0} \tilde{v}
    \sqrt{k_{x}^{2}+k_{y}^2},
\end{align}
where $\tilde{v}=v/\mu$. Restricted to the $k_z=0$ plane we recover
the two-dimensional theory, with a uniform gap
$\Delta_0\sqrt{1-\tilde{m}^2}$.  

The derivation of the momentum-averaged Green's function $G_0(\omega)$ is quite similar to
that in Appendix~\ref{sec:Appendix Green's function}, except that
accounting for the variation of the gap over the Fermi surface gives a
more complicated 
frequency-dependence:
\begin{widetext}
\begin{align}
G_{0}(\omega)=&  -\frac{N_0 \pi\tilde{\omega}}{4} \left[ \pi +i \log\left(\frac{1+\tilde{\omega}}{1-\tilde{\omega}}\right) 
  \right]\left(\eta_0  +\tilde{m}\eta_x
  \right)\sigma_0\tau_0 -  y (\eta_0 + \tilde{m} \eta_x)\sigma_0\tau_z \notag \\
& + \left[i N_0 \pi\frac{ \sqrt{1-\tilde{m}^2}  \left(\tilde{\omega}-(1+\tilde{\omega}^2) 
\tanh ^{-1}(\tilde{\omega})\right)}{4}-N_0 \pi^2 \frac{ \sqrt{1-\tilde{m}^2}  \left(
1 +\tilde{\omega}^2\right)}{8}\right] \eta_z\sigma_y \tau_y.
\label{eq:General_G}\end{align}
\end{widetext}
The virtual bound states are most clearly resolved by examining the
local density of states (LDOS) at the impurity site; the deviation of
the LDOS from the background contribution from the bulk electronic structure is given by 
\begin{equation}\label{eq:deltaLDOS}
  \delta N_{\mathrm{imp}}(\omega)=-\frac{1}{2\pi}
  \operatorname{Im}\text{Tr}\left[G_0(\omega)T(\omega)G_0(\omega) (\tau_0 + \tau_z)\right],
\end{equation}
where $T(\omega)$ is the $T$-matrix, and the factor of
$\frac{1}{2}(\tau_0+\tau_z)$ in the trace selects the electron-like
components of the Green's function.

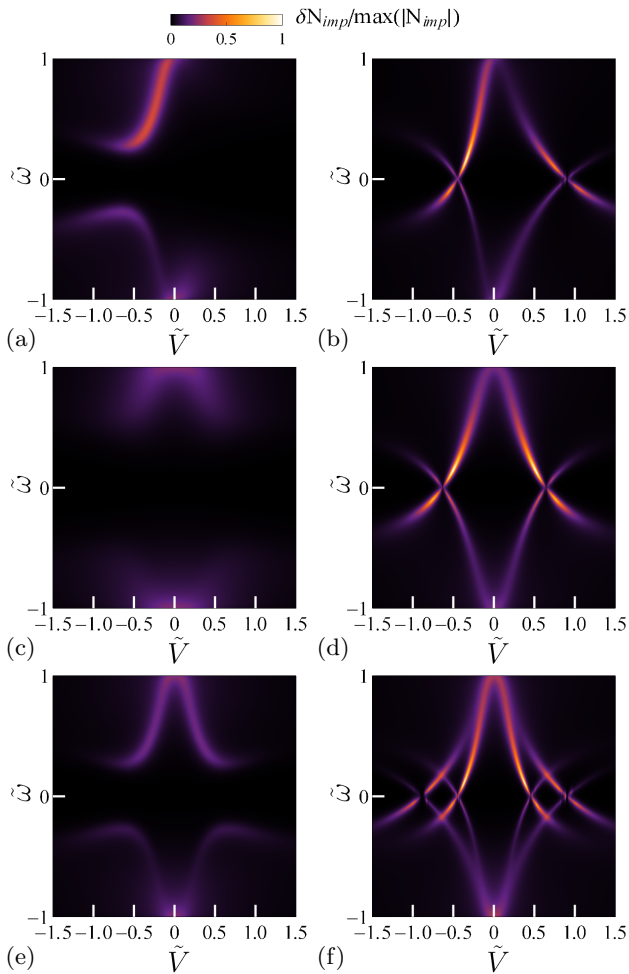
\begin{figure}[]
\centering
\vspace{-0.10in}
\hspace{-0.16in}
  \begin{overpic}[width=1.03\columnwidth, keepaspectratio]{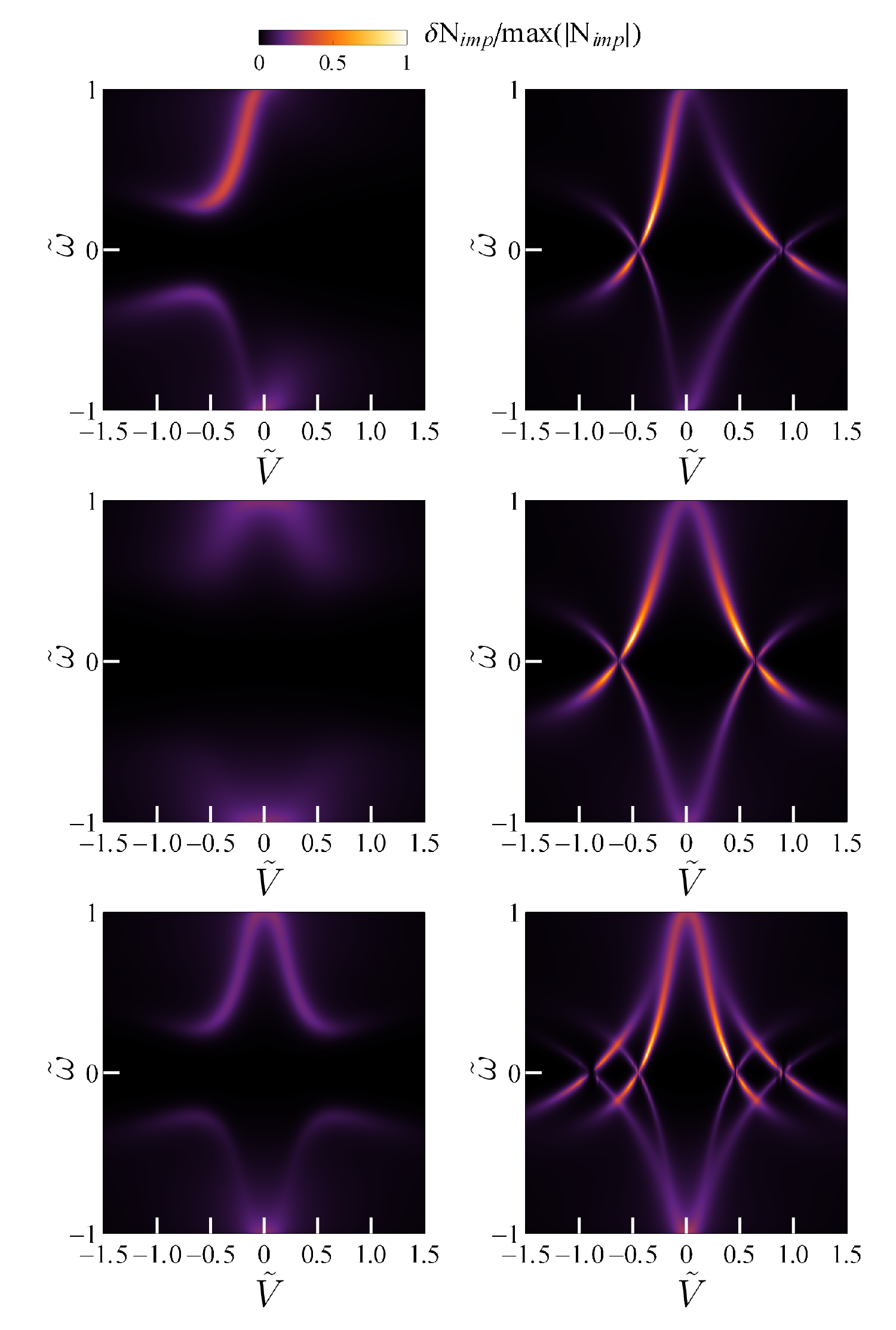}
    \put(3,  64.5){(a)}
    \put(34, 64.5){(b)}
    \put(3, 33.5){(c)}
    \put(34, 33.5){(d)}
    \put(3, 2.5){(e)}
    \put(34, 2.5){(f)}
  \end{overpic}
\vspace{-0.in}
\caption{Deviation of the LDOS at the impurity site from the bulk DOS as a function of
  dimensionless impurity strength $\tilde{V}$ for the
  $A_{\text{2u}}$ pairing state.  Panels (a)-(f) correspond to the six possible impurity
  classes summarized in Table \ref{tab:imp_survey}. 
We set $\tilde{m}=0.54$ and $\tilde{y}=0.50$.} 
\label{fig:LDOS}
\end{figure}

In Fig. \ref{fig:LDOS} we plot the deviation of the LDOS
Eq.~\ref{eq:deltaLDOS} for impurity potentials corresponding to the same classification
as in Fig.~\ref{fig:evals_vs_Vimp}. We immediately observe that the
spectrum is 
very similar to the fully-gapped two-dimensional model, but
with the impurity bound states broadened into virtual bound states due
to the hybridization with bulk quasiparticles. This indicates that the
impurity physics is still largely controlled by the fitness parameters as
found above. Case (c) appears to be an exception to this rule, as we
observe subgap features in the three-dimensional system whereas there
are no bound states in two dimensions. This discrepancy is expected
since the cancellation in Eq.~\ref{eq:tau} no longer holds: specifically the
Fermi-surface average of the fitness is now less than
$1-\tilde{t}_\perp^2$ due to the nodal gap, and so there is a finite
effective scattering rate, and thus the impurity is pair-breaking in
three dimensions. 

A number of other features of the LDOS are notable. The subgap resonances
become sharper as one approaches the middle of the gap, reflecting the
reduced density of bulk quasiparticle states, which vanish at zero
energy. The LDOS also has a pronounced particle-hole asymmetry which
arises from the particle-hole asymmetry of the normal-state~\cite{Kariyado2010,Beaird2012}. We also observe that the LDOS features fade out with
increasing impurity potential strength. This is likely due to the
shift of wavefunction weight away from the impurity site as the
potential increases. 

\subsection{$E_{\text{u}}$ pairing states}

The gaps opened by the $E_{\text{u}}$ pairing states in the three-dimensional
model have a similar form compared to the $A_{\text{2u}}$ state, albeit with
point nodes along the $x$- and $y$-axes. Indeed, the
momentum-space BdG Hamiltonian for the clean $A_{\text{2u}}$ state
can be mapped to that for each of the $E_{\text{u}}$ states; specifically, for the two
$E_{\text{u}}$ states we have
\begin{align}
{\cal H}_{\text{BdG},E_{\text{u,x}}}(k_x,k_y,k_z) = & U^\dagger_x{\cal
                                                   H}_{\text{BdG},A_{\text{2u}}}(k_z,k_y,-k_x)U_x \,, \label{eq:A2utoEux}\\
{\cal H}_{\text{BdG},E_{\text{u,y}}}(k_x,k_y,k_z) = & U^\dagger_y{\cal
                                                   H}_{\text{BdG},A_{\text{2u}}}(k_x,k_z,-k_y)U_y \,, \label{eq:A2utoEuy}\end{align}
where 
\begin{align}
U_x = & \begin{pmatrix}
\exp(-i\frac{\pi}{4}\eta_x\sigma_y) & 0 \\
0 & \exp(i\frac{\pi}{4}\eta_x\sigma_y) 
\end{pmatrix} \,, \\
U_y = & \begin{pmatrix}
\exp(i\frac{\pi}{4}\eta_x\sigma_x) & 0 \\
0 & -\exp(-i\frac{\pi}{4}\eta_x\sigma_x)
\end{pmatrix} \,.
\end{align}
Note that the swapping of the momentum components in
Eqs.~\ref{eq:A2utoEux} and~\ref{eq:A2utoEuy} does not
change the momentum-averaged Green's function, and thus does affect our
results in the $T$-matrix theory. 
Thus, as for the $A_{\text{1u}}$ and $A_{\text{2u}}$ states in the two-dimensional 
model, our results for the $A_{\text{2u}}$ state in the three-dimensional model can be mapped to each $E_{\text{u}}$ state individually, with due alteration of the impurity potential by the unitary transform. The impurity potentials corresponding to each of the subgap spectra in Fig.~\ref{fig:LDOS} are listed in Tab.~\ref{tab:imp_survey}.

\section{\label{sec:discussion}Discussion}

We have studied the appearance of impurity bound states in a model of
a superconducting bilayer using numerical self-consistent mean-field
and analytic non-self-consistent $T$-matrix methods. We find that the spectrum of bound states around an impurity is controlled by the
fitness of the pairing state with respect to both the normal-state band 
structure and the impurity potential. The bound-state spectrum due to
different impurities falls into distinct categories which are summarized 
in Table~\ref{tab:imp_survey}. 
Our results reflect the conclusion of a generalized Anderson's theorem
for the odd-parity states, namely that the pair-breaking impurities
are those which are unfit~\cite{Andersen2020,Cavanagh2021}.  
In particular, only the bound states of unfit impurities cross zero
energy with increasing impurity potential strength.  
Unlike the original Anderson's theorem, however, a finite concentration of fit impurities still typically leads to a suppression
of the critical temperature. 
Bound states can occur for these fit impurities if the pairing state is unfit with respect to the normal-state Hamiltonian, but do
not cross zero energy. 
The classification scheme also holds for
virtual bound states in the case of nodal pairing, and the subgap
spectrum shows close qualitative agreement with the true bound
states. We have thus seen how the interplay of the two types of
fitness controls the form of the bound state spectrum. The good
quantitative agreement between the numerical and analytic techniques
ensures the validity of our conclusions.

Fitness is fundamentally a way of quantifying multiband effects in a
superconductor. Our results therefore imply that multiband physics can
play an important role in superconductivity, even when
only a single band crosses the Fermi energy. It is instructive to
compare our results with Refs.~\cite{Wang2012,Mashkoori2017}, where
impurity bound states in a single-band fully-gapped unconventional
superconductor were studied for a momentum-independent impurity
potential. These works found much less diversity of the impurity bound
state spectrum. 
This apparently contradicts our findings, as the odd-parity 
$s$-wave states we study are equivalent to $p$-wave pseudospin-triplet
pairing states when projected onto the Fermi surface~\cite{Yip2013},
and thus have the same form as the states studied in
Refs.~\cite{Wang2012,Mashkoori2017}. However, the projected impurity
potential is typically both momentum- and
pseudospin-dependent~\cite{Michaeli2012,Dentelski2020}, accounting for
the difference with Refs.~\cite{Wang2012,Mashkoori2017}. The
conclusions of effective single-band models with simple impurity
potentials must therefore be treated with caution for materials where
spin and other degrees of freedom (e.g. orbital or sublattice) are
strongly mixed in the normal-state bands.

This highlights a tension in our theoretical treatment: On the one hand, the impurity Hamiltonian is more naturally expressed in
the local layer-spin basis; On the other hand, superconductivity is
more usually expressed in the band-pseudospin basis. Although
computationally convenient, the relevance of the 
odd-parity $s$-wave states to realistic materials is unclear. As
pointed out in Ref.~\cite{Cavanagh2020}, however, unconventional
disorder effects can 
be present for an arbitrary odd-parity state according to the degree
to which it resembles an odd-parity $s$-wave states when projected onto the
Fermi surface. This resemblance should manifest in the presence of the
anomalous term in the momentum-integrated Green's function Eq.~\ref{eq:Greensfunction}, albeit
with a different prefactor. As such, we expect that the impurity bound
states should obey the same classification scheme found here, although
the detailed form of the bound states may be different. 

The bilayer superconductor model considered here belongs a large class of two-band Hamiltonians which have been proposed to describe a diverse variety of compounds~\cite{FuBerg2010,Kobayashi2015,Hashimoto2016,Yanase2016,Xie_2020,Shishidou_UTe2_2021}. Our theory can be readily applied to these systems due to the similar forms of the Green's functions. Moreover, the unconventional $s$-wave states at the heart of our theory occur in any system where the electrons have additional (non-spin) degrees of freedom, and so we expect that similar impurity physics will also be present in systems with three or more bands. Directly applying our results to these cases is difficult, however, as both the relation between the gap and the fitness function Eq.~\ref{eq:effgap} and the definition of the impurity fitness Eq.~\ref{eq:genAnderson} hold only for two-band systems. Nevertheless, the concept of the fitness is valid for an arbitrary number of bands, and this should allow the development of a generalized Anderson's theorem also to more complicated systems. 

Finally, we note that the role of fitness in determining the spectrum
of the bound states suggests a general principle guiding the existence
of subgap states in a superconductor. 
It has recently been pointed out that the appearance of impurity bound
states at a magnetic impurity in an $s$-wave superconductor is 
connected to the appearance of odd-frequency pair correlations which
are localized about the impurity~\cite{Perrin2020,Suzuki2022}.
Intriguingly, the existence of odd-frequency pair correlations is
directly connected to superconducting fitness~\cite{Triola2020}. We
speculate that the conclusions of these works could equally well be
formulated in terms of fitness, and that this principle can 
hence be extended to the existence of subgap states around any
inhomogeneity in a superconductor.

\begin{acknowledgments}

Y.Z. and P.M.R.B were supported by the Marsden Fund Council from Government funding, managed by Royal
Society Te Ap\={a}rangi, Contract No. UOO1836. The authors acknowledge useful discussions
with D. C. Cavanagh, J. Schmalian, S. Rachel, A. Ramires, B. Zinkl,
and J.-X. Zhu. 

Y.Z. and N.A.H. contributed equally to this work.

\end{acknowledgments}

\bibliography{apssamp}

\begin{thebibliography}{48}%
\makeatletter
\providecommand \@ifxundefined [1]{%
 \@ifx{#1\undefined}
}%
\providecommand \@ifnum [1]{%
 \ifnum #1\expandafter \@firstoftwo
 \else \expandafter \@secondoftwo
 \fi
}%
\providecommand \@ifx [1]{%
 \ifx #1\expandafter \@firstoftwo
 \else \expandafter \@secondoftwo
 \fi
}%
\providecommand \natexlab [1]{#1}%
\providecommand \enquote  [1]{``#1''}%
\providecommand \bibnamefont  [1]{#1}%
\providecommand \bibfnamefont [1]{#1}%
\providecommand \citenamefont [1]{#1}%
\providecommand \href@noop [0]{\@secondoftwo}%
\providecommand \href [0]{\begingroup \@sanitize@url \@href}%
\providecommand \@href[1]{\@@startlink{#1}\@@href}%
\providecommand \@@href[1]{\endgroup#1\@@endlink}%
\providecommand \@sanitize@url [0]{\catcode `\\12\catcode `\$12\catcode
  `\&12\catcode `\#12\catcode `\^12\catcode `\_12\catcode `\%12\relax}%
\providecommand \@@startlink[1]{}%
\providecommand \@@endlink[0]{}%
\providecommand \url  [0]{\begingroup\@sanitize@url \@url }%
\providecommand \@url [1]{\endgroup\@href {#1}{\urlprefix }}%
\providecommand \urlprefix  [0]{URL }%
\providecommand \Eprint [0]{\href }%
\providecommand \doibase [0]{http://dx.doi.org/}%
\providecommand \selectlanguage [0]{\@gobble}%
\providecommand \bibinfo  [0]{\@secondoftwo}%
\providecommand \bibfield  [0]{\@secondoftwo}%
\providecommand \translation [1]{[#1]}%
\providecommand \BibitemOpen [0]{}%
\providecommand \bibitemStop [0]{}%
\providecommand \bibitemNoStop [0]{.\EOS\space}%
\providecommand \EOS [0]{\spacefactor3000\relax}%
\providecommand \BibitemShut  [1]{\csname bibitem#1\endcsname}%
\let\auto@bib@innerbib\@empty
%</preamble>
\bibitem [{\citenamefont {Anderson}(1959)}]{Anderson}%
  \BibitemOpen
  \bibfield  {author} {\bibinfo {author} {\bibfnamefont {P.~W.}\ \bibnamefont
  {Anderson}},\ }\bibfield  {title} {\enquote {\bibinfo {title} {Theory of
  dirty superconductors},}\ }\href {\doibase
  https://doi.org/10.1016/0022-3697(59)90036-8} {\bibfield  {journal} {\bibinfo
   {journal} {Journal of Physics and Chemistry of Solids}\ }\textbf {\bibinfo
  {volume} {11}},\ \bibinfo {pages} {26--30} (\bibinfo {year}
  {1959})}\BibitemShut {NoStop}%
\bibitem [{\citenamefont {Mineev}\ and\ \citenamefont
  {Samokhin}(1999)}]{Mineev1999}%
  \BibitemOpen
  \bibfield  {author} {\bibinfo {author} {\bibfnamefont {V.~P.}\ \bibnamefont
  {Mineev}}\ and\ \bibinfo {author} {\bibfnamefont {K.~V.}\ \bibnamefont
  {Samokhin}},\ }\href@noop {} {\emph {\bibinfo {title} {{Introduction to
  Unconventional Superconductivity}}}}\ (\bibinfo  {publisher} {Gordon and
  Breach Science Publishers},\ \bibinfo {year} {1999})\BibitemShut {NoStop}%
\bibitem [{\citenamefont {Yu}(1965)}]{Yu1965}%
  \BibitemOpen
  \bibfield  {author} {\bibinfo {author} {\bibfnamefont {L.}~\bibnamefont
  {Yu}},\ }\bibfield  {title} {\enquote {\bibinfo {title} {Bound state in
  superconductors with paramagnetic impurities},}\ }\href {\doibase
  10.7498/aps.21.75} {\bibfield  {journal} {\bibinfo  {journal} {Acta Phys.
  Sin.}\ }\textbf {\bibinfo {volume} {21}},\ \bibinfo {pages} {75} (\bibinfo
  {year} {1965})}\BibitemShut {NoStop}%
\bibitem [{\citenamefont {Shiba}(1968)}]{Shiba1968}%
  \BibitemOpen
  \bibfield  {author} {\bibinfo {author} {\bibfnamefont {H.}~\bibnamefont
  {Shiba}},\ }\bibfield  {title} {\enquote {\bibinfo {title} {{Classical Spins
  in Superconductors}},}\ }\href {\doibase 10.1143/PTP.40.435} {\bibfield
  {journal} {\bibinfo  {journal} {Prog. Theor. Phys.}\ }\textbf {\bibinfo
  {volume} {40}},\ \bibinfo {pages} {435} (\bibinfo {year} {1968})}\BibitemShut
  {NoStop}%
\bibitem [{\citenamefont {Rusinov}(1969)}]{Rusinov1969}%
  \BibitemOpen
  \bibfield  {author} {\bibinfo {author} {\bibfnamefont {A.~I.}\ \bibnamefont
  {Rusinov}},\ }\bibfield  {title} {\enquote {\bibinfo {title} {{On the Theory
  of Gapless Superconductivity in Alloys Containing Paramagnetic
  Impurities}},}\ }\href@noop {} {\bibfield  {journal} {\bibinfo  {journal}
  {Sov. J. Exp. Theor. Phys.}\ }\textbf {\bibinfo {volume} {29}},\ \bibinfo
  {pages} {1101} (\bibinfo {year} {1969})}\BibitemShut {NoStop}%
\bibitem [{\citenamefont {Wang}\ and\ \citenamefont {Wang}(2004)}]{Wang2004}%
  \BibitemOpen
  \bibfield  {author} {\bibinfo {author} {\bibfnamefont {Q.-H.}\ \bibnamefont
  {Wang}}\ and\ \bibinfo {author} {\bibfnamefont {Z.~D.}\ \bibnamefont
  {Wang}},\ }\bibfield  {title} {\enquote {\bibinfo {title} {{Impurity and
  interface bound states in
  ${d}_{{x}^{2}\ensuremath{-}{y}^{2}}{+id}_{\mathrm{xy}}$ and
  ${p}_{x}{+ip}_{y}$ superconductors}},}\ }\href {\doibase
  10.1103/PhysRevB.69.092502} {\bibfield  {journal} {\bibinfo  {journal} {Phys.
  Rev. B}\ }\textbf {\bibinfo {volume} {69}},\ \bibinfo {pages} {092502}
  (\bibinfo {year} {2004})}\BibitemShut {NoStop}%
\bibitem [{\citenamefont {Wang}\ \emph {et~al.}(2012)\citenamefont {Wang},
  \citenamefont {Liu}, \citenamefont {Ma},\ and\ \citenamefont
  {Jiang}}]{Wang2012}%
  \BibitemOpen
  \bibfield  {author} {\bibinfo {author} {\bibfnamefont {F.}~\bibnamefont
  {Wang}}, \bibinfo {author} {\bibfnamefont {Q.}~\bibnamefont {Liu}}, \bibinfo
  {author} {\bibfnamefont {T.}~\bibnamefont {Ma}}, \ and\ \bibinfo {author}
  {\bibfnamefont {X.}~\bibnamefont {Jiang}},\ }\bibfield  {title} {\enquote
  {\bibinfo {title} {Impurity-induced bound states in superconductors with
  topological order},}\ }\href {\doibase 10.1088/0953-8984/24/45/455701}
  {\bibfield  {journal} {\bibinfo  {journal} {Journal of Physics Condensed
  Matter}\ }\textbf {\bibinfo {volume} {24}},\ \bibinfo {pages} {455701}
  (\bibinfo {year} {2012})}\BibitemShut {NoStop}%
\bibitem [{\citenamefont {Kim}\ \emph {et~al.}(2015)\citenamefont {Kim},
  \citenamefont {Zhang}, \citenamefont {Rossi},\ and\ \citenamefont
  {Lutchyn}}]{Kim2015}%
  \BibitemOpen
  \bibfield  {author} {\bibinfo {author} {\bibfnamefont {Y.}~\bibnamefont
  {Kim}}, \bibinfo {author} {\bibfnamefont {J.}~\bibnamefont {Zhang}}, \bibinfo
  {author} {\bibfnamefont {E.}~\bibnamefont {Rossi}}, \ and\ \bibinfo {author}
  {\bibfnamefont {R.~M.}\ \bibnamefont {Lutchyn}},\ }\bibfield  {title}
  {\enquote {\bibinfo {title} {{Impurity-Induced Bound States in
  Superconductors with Spin-Orbit Coupling}},}\ }\href {\doibase
  10.1103/PhysRevLett.114.236804} {\bibfield  {journal} {\bibinfo  {journal}
  {Phys. Rev. Lett.}\ }\textbf {\bibinfo {volume} {114}},\ \bibinfo {pages}
  {236804} (\bibinfo {year} {2015})}\BibitemShut {NoStop}%
\bibitem [{\citenamefont {Mashkoori}\ \emph {et~al.}(2017)\citenamefont
  {Mashkoori}, \citenamefont {Bj{\"o}rnson},\ and\ \citenamefont
  {Black-Schaffer}}]{Mashkoori2017}%
  \BibitemOpen
  \bibfield  {author} {\bibinfo {author} {\bibfnamefont {M.}~\bibnamefont
  {Mashkoori}}, \bibinfo {author} {\bibfnamefont {K.}~\bibnamefont
  {Bj{\"o}rnson}}, \ and\ \bibinfo {author} {\bibfnamefont {A.~M.}\
  \bibnamefont {Black-Schaffer}},\ }\bibfield  {title} {\enquote {\bibinfo
  {title} {{Impurity bound states in fully gapped $d$-wave superconductors with
  subdominant order parameters}},}\ }\href {\doibase
  https://doi.org/10.1038/srep44107} {\bibfield  {journal} {\bibinfo  {journal}
  {Scientific Reports}\ }\textbf {\bibinfo {volume} {7}},\ \bibinfo {pages}
  {44107} (\bibinfo {year} {2017})}\BibitemShut {NoStop}%
\bibitem [{\citenamefont {Balatsky}\ \emph {et~al.}(1995)\citenamefont
  {Balatsky}, \citenamefont {Salkola},\ and\ \citenamefont
  {Rosengren}}]{Balatsky1995}%
  \BibitemOpen
  \bibfield  {author} {\bibinfo {author} {\bibfnamefont {A.~V.}\ \bibnamefont
  {Balatsky}}, \bibinfo {author} {\bibfnamefont {M.~I.}\ \bibnamefont
  {Salkola}}, \ and\ \bibinfo {author} {\bibfnamefont {A.}~\bibnamefont
  {Rosengren}},\ }\bibfield  {title} {\enquote {\bibinfo {title}
  {{Impurity-induced virtual bound states in $d$-wave superconductors}},}\
  }\href {\doibase 10.1103/PhysRevB.51.15547} {\bibfield  {journal} {\bibinfo
  {journal} {Phys. Rev. B}\ }\textbf {\bibinfo {volume} {51}},\ \bibinfo
  {pages} {15547} (\bibinfo {year} {1995})}\BibitemShut {NoStop}%
\bibitem [{\citenamefont {Liu}\ and\ \citenamefont {Eremin}(2008)}]{Liu2008}%
  \BibitemOpen
  \bibfield  {author} {\bibinfo {author} {\bibfnamefont {B.}~\bibnamefont
  {Liu}}\ and\ \bibinfo {author} {\bibfnamefont {I.}~\bibnamefont {Eremin}},\
  }\bibfield  {title} {\enquote {\bibinfo {title} {{Impurity resonance states
  in noncentrosymmetric superconductor ${\text{CePt}}_{3}\text{Si}$: A probe
  for Cooper-pairing symmetry}},}\ }\href {\doibase 10.1103/PhysRevB.78.014518}
  {\bibfield  {journal} {\bibinfo  {journal} {Phys. Rev. B}\ }\textbf {\bibinfo
  {volume} {78}},\ \bibinfo {pages} {014518} (\bibinfo {year}
  {2008})}\BibitemShut {NoStop}%
\bibitem [{\citenamefont {Balatsky}\ \emph {et~al.}(2006)\citenamefont
  {Balatsky}, \citenamefont {Vekhter},\ and\ \citenamefont
  {Zhu}}]{BalatskyRMP2006}%
  \BibitemOpen
  \bibfield  {author} {\bibinfo {author} {\bibfnamefont {A.~V.}\ \bibnamefont
  {Balatsky}}, \bibinfo {author} {\bibfnamefont {I.}~\bibnamefont {Vekhter}}, \
  and\ \bibinfo {author} {\bibfnamefont {J.-X.}\ \bibnamefont {Zhu}},\
  }\bibfield  {title} {\enquote {\bibinfo {title} {Impurity-induced states in
  conventional and unconventional superconductors},}\ }\href {\doibase
  10.1103/RevModPhys.78.373} {\bibfield  {journal} {\bibinfo  {journal} {Rev.
  Mod. Phys.}\ }\textbf {\bibinfo {volume} {78}},\ \bibinfo {pages} {373--433}
  (\bibinfo {year} {2006})}\BibitemShut {NoStop}%
\bibitem [{\citenamefont {Yazdani}\ \emph {et~al.}(1997)\citenamefont
  {Yazdani}, \citenamefont {Jones}, \citenamefont {Lutz}, \citenamefont
  {Crommie},\ and\ \citenamefont {Eigler}}]{Yazdani1997}%
  \BibitemOpen
  \bibfield  {author} {\bibinfo {author} {\bibfnamefont {A.}~\bibnamefont
  {Yazdani}}, \bibinfo {author} {\bibfnamefont {B.~A.}\ \bibnamefont {Jones}},
  \bibinfo {author} {\bibfnamefont {C.~P.}\ \bibnamefont {Lutz}}, \bibinfo
  {author} {\bibfnamefont {M.~F.}\ \bibnamefont {Crommie}}, \ and\ \bibinfo
  {author} {\bibfnamefont {D.~M.}\ \bibnamefont {Eigler}},\ }\bibfield  {title}
  {\enquote {\bibinfo {title} {{Probing the Local Effects of Magnetic
  Impurities on Superconductivity}},}\ }\href {\doibase
  10.1126/science.275.5307.1767} {\bibfield  {journal} {\bibinfo  {journal}
  {Science}\ }\textbf {\bibinfo {volume} {275}},\ \bibinfo {pages} {1767}
  (\bibinfo {year} {1997})}\BibitemShut {NoStop}%
\bibitem [{\citenamefont {Hudson}\ \emph {et~al.}(2001)\citenamefont {Hudson},
  \citenamefont {Lang}, \citenamefont {Madhavan}, \citenamefont {Pan},
  \citenamefont {Eisaki}, \citenamefont {Uchida},\ and\ \citenamefont
  {Davis}}]{Hudson2001}%
  \BibitemOpen
  \bibfield  {author} {\bibinfo {author} {\bibfnamefont {E.~W.}\ \bibnamefont
  {Hudson}}, \bibinfo {author} {\bibfnamefont {K.~M.}\ \bibnamefont {Lang}},
  \bibinfo {author} {\bibfnamefont {V.}~\bibnamefont {Madhavan}}, \bibinfo
  {author} {\bibfnamefont {S.~H.}\ \bibnamefont {Pan}}, \bibinfo {author}
  {\bibfnamefont {H.}~\bibnamefont {Eisaki}}, \bibinfo {author} {\bibfnamefont
  {S.}~\bibnamefont {Uchida}}, \ and\ \bibinfo {author} {\bibfnamefont {J.~C.}\
  \bibnamefont {Davis}},\ }\bibfield  {title} {\enquote {\bibinfo {title}
  {{Interplay of magnetism and high-Tc superconductivity at individual Ni
  impurity atoms in Bi$_2$Sr$_2$CaCu$_2$O$_{8+\delta}$}},}\ }\href {\doibase
  https://doi.org/10.1038/35082019} {\bibfield  {journal} {\bibinfo  {journal}
  {Nature}\ }\textbf {\bibinfo {volume} {411}},\ \bibinfo {pages} {920–924}
  (\bibinfo {year} {2001})}\BibitemShut {NoStop}%
\bibitem [{\citenamefont {Grothe}\ \emph {et~al.}(2012)\citenamefont {Grothe},
  \citenamefont {Chi}, \citenamefont {Dosanjh}, \citenamefont {Liang},
  \citenamefont {Hardy}, \citenamefont {Burke}, \citenamefont {Bonn},\ and\
  \citenamefont {Pennec}}]{Grothe2012}%
  \BibitemOpen
  \bibfield  {author} {\bibinfo {author} {\bibfnamefont {S.}~\bibnamefont
  {Grothe}}, \bibinfo {author} {\bibfnamefont {S.}~\bibnamefont {Chi}},
  \bibinfo {author} {\bibfnamefont {P.}~\bibnamefont {Dosanjh}}, \bibinfo
  {author} {\bibfnamefont {R.}~\bibnamefont {Liang}}, \bibinfo {author}
  {\bibfnamefont {W.~N.}\ \bibnamefont {Hardy}}, \bibinfo {author}
  {\bibfnamefont {S.~A.}\ \bibnamefont {Burke}}, \bibinfo {author}
  {\bibfnamefont {D.~A.}\ \bibnamefont {Bonn}}, \ and\ \bibinfo {author}
  {\bibfnamefont {Y.}~\bibnamefont {Pennec}},\ }\bibfield  {title} {\enquote
  {\bibinfo {title} {{Bound states of defects in superconducting LiFeAs studied
  by scanning tunneling spectroscopy}},}\ }\href {\doibase
  10.1103/PhysRevB.86.174503} {\bibfield  {journal} {\bibinfo  {journal} {Phys.
  Rev. B}\ }\textbf {\bibinfo {volume} {86}},\ \bibinfo {pages} {174503}
  (\bibinfo {year} {2012})}\BibitemShut {NoStop}%
\bibitem [{\citenamefont {Nadj-Perge}\ \emph {et~al.}(2014)\citenamefont
  {Nadj-Perge}, \citenamefont {Drozdov}, \citenamefont {Li}, \citenamefont
  {Chen}, \citenamefont {Jeon}, \citenamefont {Seo}, \citenamefont {MacDonald},
  \citenamefont {Bernevig},\ and\ \citenamefont {Yazdani}}]{NadjPerge2014}%
  \BibitemOpen
  \bibfield  {author} {\bibinfo {author} {\bibfnamefont {S.}~\bibnamefont
  {Nadj-Perge}}, \bibinfo {author} {\bibfnamefont {I.~K.}\ \bibnamefont
  {Drozdov}}, \bibinfo {author} {\bibfnamefont {J.}~\bibnamefont {Li}},
  \bibinfo {author} {\bibfnamefont {H.}~\bibnamefont {Chen}}, \bibinfo {author}
  {\bibfnamefont {S.}~\bibnamefont {Jeon}}, \bibinfo {author} {\bibfnamefont
  {J.}~\bibnamefont {Seo}}, \bibinfo {author} {\bibfnamefont {A.~H.}\
  \bibnamefont {MacDonald}}, \bibinfo {author} {\bibfnamefont {B.~A.}\
  \bibnamefont {Bernevig}}, \ and\ \bibinfo {author} {\bibfnamefont
  {A.}~\bibnamefont {Yazdani}},\ }\bibfield  {title} {\enquote {\bibinfo
  {title} {{Observation of Majorana fermions in ferromagnetic atomic chains on
  a superconductor}},}\ }\href {\doibase 10.1126/science.1259327} {\bibfield
  {journal} {\bibinfo  {journal} {Science}\ }\textbf {\bibinfo {volume}
  {346}},\ \bibinfo {pages} {602} (\bibinfo {year} {2014})}\BibitemShut
  {NoStop}%
\bibitem [{\citenamefont {Pawlak}\ \emph {et~al.}(2016)\citenamefont {Pawlak},
  \citenamefont {Kisiel}, \citenamefont {Klinovaja}, \citenamefont {Meier},
  \citenamefont {Kawai}, \citenamefont {Glatzel}, \citenamefont {Loss},\ and\
  \citenamefont {Meyer}}]{Pawlak2016}%
  \BibitemOpen
  \bibfield  {author} {\bibinfo {author} {\bibfnamefont {R.}~\bibnamefont
  {Pawlak}}, \bibinfo {author} {\bibfnamefont {M.}~\bibnamefont {Kisiel}},
  \bibinfo {author} {\bibfnamefont {J.}~\bibnamefont {Klinovaja}}, \bibinfo
  {author} {\bibfnamefont {T.}~\bibnamefont {Meier}}, \bibinfo {author}
  {\bibfnamefont {S.}~\bibnamefont {Kawai}}, \bibinfo {author} {\bibfnamefont
  {T.}~\bibnamefont {Glatzel}}, \bibinfo {author} {\bibfnamefont
  {D.}~\bibnamefont {Loss}}, \ and\ \bibinfo {author} {\bibfnamefont
  {E.}~\bibnamefont {Meyer}},\ }\bibfield  {title} {\enquote {\bibinfo {title}
  {{Probing atomic structure and Majorana wavefunctions in mono-atomic Fe
  chains on superconducting Pb surface}},}\ }\href {\doibase
  https://doi.org/10.1038/npjqi.2016.35} {\bibfield  {journal} {\bibinfo
  {journal} {npj Quantum Information}\ }\textbf {\bibinfo {volume} {2}},\
  \bibinfo {pages} {16035} (\bibinfo {year} {2016})}\BibitemShut {NoStop}%
\bibitem [{\citenamefont {Schneider}\ \emph {et~al.}(2021)\citenamefont
  {Schneider}, \citenamefont {Beck}, \citenamefont {Posske}, \citenamefont
  {Crawford}, \citenamefont {Mascot}, \citenamefont {Rachel}, \citenamefont
  {Wiesendanger},\ and\ \citenamefont {Wiebe}}]{Schneider2021}%
  \BibitemOpen
  \bibfield  {author} {\bibinfo {author} {\bibfnamefont {L.}~\bibnamefont
  {Schneider}}, \bibinfo {author} {\bibfnamefont {P.}~\bibnamefont {Beck}},
  \bibinfo {author} {\bibfnamefont {T.}~\bibnamefont {Posske}}, \bibinfo
  {author} {\bibfnamefont {D.}~\bibnamefont {Crawford}}, \bibinfo {author}
  {\bibfnamefont {E.}~\bibnamefont {Mascot}}, \bibinfo {author} {\bibfnamefont
  {S.}~\bibnamefont {Rachel}}, \bibinfo {author} {\bibfnamefont
  {R.}~\bibnamefont {Wiesendanger}}, \ and\ \bibinfo {author} {\bibfnamefont
  {J.}~\bibnamefont {Wiebe}},\ }\bibfield  {title} {\enquote {\bibinfo {title}
  {{Topological Shiba bands in artificial spin chains on superconductors}},}\
  }\href {\doibase https://doi.org/10.1038/s41567-021-01234-y} {\bibfield
  {journal} {\bibinfo  {journal} {Nature Physics}\ }\textbf {\bibinfo {volume}
  {17}},\ \bibinfo {pages} {943–948} (\bibinfo {year} {2021})}\BibitemShut
  {NoStop}%
\bibitem [{\citenamefont {Kriener}\ \emph {et~al.}(2012)\citenamefont
  {Kriener}, \citenamefont {Segawa}, \citenamefont {Sasaki},\ and\
  \citenamefont {Ando}}]{Kriener2012}%
  \BibitemOpen
  \bibfield  {author} {\bibinfo {author} {\bibfnamefont {M.}~\bibnamefont
  {Kriener}}, \bibinfo {author} {\bibfnamefont {K.}~\bibnamefont {Segawa}},
  \bibinfo {author} {\bibfnamefont {S.}~\bibnamefont {Sasaki}}, \ and\ \bibinfo
  {author} {\bibfnamefont {Y.}~\bibnamefont {Ando}},\ }\bibfield  {title}
  {\enquote {\bibinfo {title} {{Anomalous suppression of the superfluid density
  in the Cu${}_{x}$Bi${}_{2}$Se${}_{3}$ superconductor upon progressive Cu
  intercalation}},}\ }\href {\doibase 10.1103/PhysRevB.86.180505} {\bibfield
  {journal} {\bibinfo  {journal} {Phys. Rev. B}\ }\textbf {\bibinfo {volume}
  {86}},\ \bibinfo {pages} {180505} (\bibinfo {year} {2012})}\BibitemShut
  {NoStop}%
\bibitem [{\citenamefont {Smylie}\ \emph {et~al.}(2017)\citenamefont {Smylie},
  \citenamefont {Willa}, \citenamefont {Claus}, \citenamefont {Snezhko},
  \citenamefont {Martin}, \citenamefont {Kwok}, \citenamefont {Qiu},
  \citenamefont {Hor}, \citenamefont {Bokari}, \citenamefont {Niraula},
  \citenamefont {Kayani}, \citenamefont {Mishra},\ and\ \citenamefont
  {Welp}}]{Smylie2017}%
  \BibitemOpen
  \bibfield  {author} {\bibinfo {author} {\bibfnamefont {M.~P.}\ \bibnamefont
  {Smylie}}, \bibinfo {author} {\bibfnamefont {K.}~\bibnamefont {Willa}},
  \bibinfo {author} {\bibfnamefont {H.}~\bibnamefont {Claus}}, \bibinfo
  {author} {\bibfnamefont {A.}~\bibnamefont {Snezhko}}, \bibinfo {author}
  {\bibfnamefont {I.}~\bibnamefont {Martin}}, \bibinfo {author} {\bibfnamefont
  {W.-K.}\ \bibnamefont {Kwok}}, \bibinfo {author} {\bibfnamefont
  {Y.}~\bibnamefont {Qiu}}, \bibinfo {author} {\bibfnamefont {Y.~S.}\
  \bibnamefont {Hor}}, \bibinfo {author} {\bibfnamefont {E.}~\bibnamefont
  {Bokari}}, \bibinfo {author} {\bibfnamefont {P.}~\bibnamefont {Niraula}},
  \bibinfo {author} {\bibfnamefont {A.}~\bibnamefont {Kayani}}, \bibinfo
  {author} {\bibfnamefont {V.}~\bibnamefont {Mishra}}, \ and\ \bibinfo {author}
  {\bibfnamefont {U.}~\bibnamefont {Welp}},\ }\bibfield  {title} {\enquote
  {\bibinfo {title} {{Robust odd-parity superconductivity in the doped
  topological insulator
  ${\mathrm{Nb}}_{x}{\mathrm{Bi}}_{2}{\mathrm{Se}}_{3}$}},}\ }\href {\doibase
  10.1103/PhysRevB.96.115145} {\bibfield  {journal} {\bibinfo  {journal} {Phys.
  Rev. B}\ }\textbf {\bibinfo {volume} {96}},\ \bibinfo {pages} {115145}
  (\bibinfo {year} {2017})}\BibitemShut {NoStop}%
\bibitem [{\citenamefont {Andersen}\ \emph {et~al.}(2020)\citenamefont
  {Andersen}, \citenamefont {Ramires}, \citenamefont {Wang}, \citenamefont
  {Lorenz},\ and\ \citenamefont {Ando}}]{Andersen2020}%
  \BibitemOpen
  \bibfield  {author} {\bibinfo {author} {\bibfnamefont {L.}~\bibnamefont
  {Andersen}}, \bibinfo {author} {\bibfnamefont {A.}~\bibnamefont {Ramires}},
  \bibinfo {author} {\bibfnamefont {Z.}~\bibnamefont {Wang}}, \bibinfo {author}
  {\bibfnamefont {T.}~\bibnamefont {Lorenz}}, \ and\ \bibinfo {author}
  {\bibfnamefont {Y.}~\bibnamefont {Ando}},\ }\bibfield  {title} {\enquote
  {\bibinfo {title} {{Generalized Anderson{\textquoteright}s theorem for
  superconductors derived from topological insulators}},}\ }\href {\doibase
  10.1126/sciadv.aay6502} {\bibfield  {journal} {\bibinfo  {journal} {Science
  Advances}\ }\textbf {\bibinfo {volume} {6}} (\bibinfo {year} {2020}),\
  10.1126/sciadv.aay6502}\BibitemShut {NoStop}%
\bibitem [{\citenamefont {Yonezawa}(2019)}]{Yonezawa2019}%
  \BibitemOpen
  \bibfield  {author} {\bibinfo {author} {\bibfnamefont {S.}~\bibnamefont
  {Yonezawa}},\ }\bibfield  {title} {\enquote {\bibinfo {title} {{Nematic
  Superconductivity in Doped Bi$_2$Se$_3$ Topological Superconductors}},}\
  }\href {\doibase 10.3390/condmat4010002} {\bibfield  {journal} {\bibinfo
  {journal} {Condensed Matter}\ }\textbf {\bibinfo {volume} {4}},\ \bibinfo
  {pages} {2} (\bibinfo {year} {2019})}\BibitemShut {NoStop}%
\bibitem [{\citenamefont {Michaeli}\ and\ \citenamefont
  {Fu}(2012)}]{Michaeli2012}%
  \BibitemOpen
  \bibfield  {author} {\bibinfo {author} {\bibfnamefont {K.}~\bibnamefont
  {Michaeli}}\ and\ \bibinfo {author} {\bibfnamefont {L.}~\bibnamefont {Fu}},\
  }\bibfield  {title} {\enquote {\bibinfo {title} {{Spin-Orbit Locking as a
  Protection Mechanism of the Odd-Parity Superconducting State against
  Disorder}},}\ }\href {\doibase 10.1103/PhysRevLett.109.187003} {\bibfield
  {journal} {\bibinfo  {journal} {Phys. Rev. Lett.}\ }\textbf {\bibinfo
  {volume} {109}},\ \bibinfo {pages} {187003} (\bibinfo {year}
  {2012})}\BibitemShut {NoStop}%
\bibitem [{\citenamefont {Nagai}\ \emph {et~al.}(2014)\citenamefont {Nagai},
  \citenamefont {Ota},\ and\ \citenamefont {Machida}}]{Nagai2014}%
  \BibitemOpen
  \bibfield  {author} {\bibinfo {author} {\bibfnamefont {Y.}~\bibnamefont
  {Nagai}}, \bibinfo {author} {\bibfnamefont {Y.}~\bibnamefont {Ota}}, \ and\
  \bibinfo {author} {\bibfnamefont {M.}~\bibnamefont {Machida}},\ }\bibfield
  {title} {\enquote {\bibinfo {title} {{Nonmagnetic impurity effects in a
  three-dimensional topological superconductor: From $p$- to $s$-wave
  behaviors}},}\ }\href {\doibase 10.1103/PhysRevB.89.214506} {\bibfield
  {journal} {\bibinfo  {journal} {Phys. Rev. B}\ }\textbf {\bibinfo {volume}
  {89}},\ \bibinfo {pages} {214506} (\bibinfo {year} {2014})}\BibitemShut
  {NoStop}%
\bibitem [{\citenamefont {Nagai}(2015)}]{Nagai2015}%
  \BibitemOpen
  \bibfield  {author} {\bibinfo {author} {\bibfnamefont {Y.}~\bibnamefont
  {Nagai}},\ }\bibfield  {title} {\enquote {\bibinfo {title} {{Robust
  superconductivity with nodes in the superconducting topological insulator
  ${\text{Cu}}_{x}{\text{Bi}}_{2}{\text{Se}}_{3}$: Zeeman orbital field and
  nonmagnetic impurities}},}\ }\href {\doibase 10.1103/PhysRevB.91.060502}
  {\bibfield  {journal} {\bibinfo  {journal} {Phys. Rev. B}\ }\textbf {\bibinfo
  {volume} {91}},\ \bibinfo {pages} {060502} (\bibinfo {year}
  {2015})}\BibitemShut {NoStop}%
\bibitem [{\citenamefont {Cavanagh}\ and\ \citenamefont
  {Brydon}(2020)}]{Cavanagh2020}%
  \BibitemOpen
  \bibfield  {author} {\bibinfo {author} {\bibfnamefont {D.~C.}\ \bibnamefont
  {Cavanagh}}\ and\ \bibinfo {author} {\bibfnamefont {P.~M.~R.}\ \bibnamefont
  {Brydon}},\ }\bibfield  {title} {\enquote {\bibinfo {title} {Robustness of
  unconventional $s$-wave superconducting states against disorder},}\ }\href
  {\doibase 10.1103/PhysRevB.101.054509} {\bibfield  {journal} {\bibinfo
  {journal} {Phys. Rev. B}\ }\textbf {\bibinfo {volume} {101}},\ \bibinfo
  {pages} {054509} (\bibinfo {year} {2020})}\BibitemShut {NoStop}%
\bibitem [{\citenamefont {Cavanagh}\ and\ \citenamefont
  {Brydon}(2021)}]{Cavanagh2021}%
  \BibitemOpen
  \bibfield  {author} {\bibinfo {author} {\bibfnamefont {D.~C.}\ \bibnamefont
  {Cavanagh}}\ and\ \bibinfo {author} {\bibfnamefont {P.~M.~R.}\ \bibnamefont
  {Brydon}},\ }\bibfield  {title} {\enquote {\bibinfo {title} {General theory
  of robustness against disorder in multiband superconductors},}\ }\href
  {\doibase 10.1103/PhysRevB.104.014503} {\bibfield  {journal} {\bibinfo
  {journal} {Phys. Rev. B}\ }\textbf {\bibinfo {volume} {104}},\ \bibinfo
  {pages} {014503} (\bibinfo {year} {2021})}\BibitemShut {NoStop}%
\bibitem [{\citenamefont {Dentelski}\ \emph {et~al.}(2020)\citenamefont
  {Dentelski}, \citenamefont {Kozii},\ and\ \citenamefont
  {Ruhman}}]{Dentelski2020}%
  \BibitemOpen
  \bibfield  {author} {\bibinfo {author} {\bibfnamefont {D.}~\bibnamefont
  {Dentelski}}, \bibinfo {author} {\bibfnamefont {V.}~\bibnamefont {Kozii}}, \
  and\ \bibinfo {author} {\bibfnamefont {J.}~\bibnamefont {Ruhman}},\
  }\bibfield  {title} {\enquote {\bibinfo {title} {Effect of interorbital
  scattering on superconductivity in doped dirac semimetals},}\ }\href
  {\doibase 10.1103/PhysRevResearch.2.033302} {\bibfield  {journal} {\bibinfo
  {journal} {Phys. Rev. Research}\ }\textbf {\bibinfo {volume} {2}},\ \bibinfo
  {pages} {033302} (\bibinfo {year} {2020})}\BibitemShut {NoStop}%
\bibitem [{\citenamefont {Sato}\ and\ \citenamefont {Asano}(2020)}]{Sato2020}%
  \BibitemOpen
  \bibfield  {author} {\bibinfo {author} {\bibfnamefont {T.}~\bibnamefont
  {Sato}}\ and\ \bibinfo {author} {\bibfnamefont {Y.}~\bibnamefont {Asano}},\
  }\bibfield  {title} {\enquote {\bibinfo {title} {{Superconductivity in
  Cu-doped ${\mathrm{Bi}}_{2}{\mathrm{Se}}_{3}$ with potential disorder}},}\
  }\href {\doibase 10.1103/PhysRevB.102.024516} {\bibfield  {journal} {\bibinfo
   {journal} {Phys. Rev. B}\ }\textbf {\bibinfo {volume} {102}},\ \bibinfo
  {pages} {024516} (\bibinfo {year} {2020})}\BibitemShut {NoStop}%
\bibitem [{\citenamefont {Fu}\ and\ \citenamefont {Berg}(2010)}]{FuBerg2010}%
  \BibitemOpen
  \bibfield  {author} {\bibinfo {author} {\bibfnamefont {L.}~\bibnamefont
  {Fu}}\ and\ \bibinfo {author} {\bibfnamefont {E.}~\bibnamefont {Berg}},\
  }\bibfield  {title} {\enquote {\bibinfo {title} {{Odd-Parity Topological
  Superconductors: Theory and Application to
  ${\mathrm{Cu}}_{x}{\mathrm{Bi}}_{2}{\mathrm{Se}}_{3}$}},}\ }\href {\doibase
  10.1103/PhysRevLett.105.097001} {\bibfield  {journal} {\bibinfo  {journal}
  {Phys. Rev. Lett.}\ }\textbf {\bibinfo {volume} {105}},\ \bibinfo {pages}
  {097001} (\bibinfo {year} {2010})}\BibitemShut {NoStop}%
\bibitem [{\citenamefont {Ramires}\ and\ \citenamefont
  {Sigrist}(2016)}]{Ramires2016}%
  \BibitemOpen
  \bibfield  {author} {\bibinfo {author} {\bibfnamefont {A.}~\bibnamefont
  {Ramires}}\ and\ \bibinfo {author} {\bibfnamefont {M.}~\bibnamefont
  {Sigrist}},\ }\bibfield  {title} {\enquote {\bibinfo {title} {{Identifying
  detrimental effects for multiorbital superconductivity: Application to
  ${\text{Sr}}_{2}{\text{RuO}}_{4}$}},}\ }\href {\doibase
  10.1103/PhysRevB.94.104501} {\bibfield  {journal} {\bibinfo  {journal} {Phys.
  Rev. B}\ }\textbf {\bibinfo {volume} {94}},\ \bibinfo {pages} {104501}
  (\bibinfo {year} {2016})}\BibitemShut {NoStop}%
\bibitem [{\citenamefont {Ramires}\ \emph {et~al.}(2018)\citenamefont
  {Ramires}, \citenamefont {Agterberg},\ and\ \citenamefont
  {Sigrist}}]{Ramires2018}%
  \BibitemOpen
  \bibfield  {author} {\bibinfo {author} {\bibfnamefont {A.}~\bibnamefont
  {Ramires}}, \bibinfo {author} {\bibfnamefont {D.~F.}\ \bibnamefont
  {Agterberg}}, \ and\ \bibinfo {author} {\bibfnamefont {M.}~\bibnamefont
  {Sigrist}},\ }\bibfield  {title} {\enquote {\bibinfo {title} {Tailoring
  $\text{T}_{c}$ by symmetry principles: The concept of superconducting
  fitness},}\ }\href {\doibase 10.1103/PhysRevB.98.024501} {\bibfield
  {journal} {\bibinfo  {journal} {Phys. Rev. B}\ }\textbf {\bibinfo {volume}
  {98}},\ \bibinfo {pages} {024501} (\bibinfo {year} {2018})}\BibitemShut
  {NoStop}%
\bibitem [{\citenamefont {Timmons}\ \emph {et~al.}(2020)\citenamefont
  {Timmons}, \citenamefont {Teknowijoyo}, \citenamefont
  {Ko\ifmmode~\acute{n}\else \'{n}\fi{}czykowski}, \citenamefont {Cavani},
  \citenamefont {Tanatar}, \citenamefont {Ghimire}, \citenamefont {Cho},
  \citenamefont {Lee}, \citenamefont {Ke}, \citenamefont {Jo}, \citenamefont
  {Bud'ko}, \citenamefont {Canfield}, \citenamefont {Orth}, \citenamefont
  {Scheurer},\ and\ \citenamefont {Prozorov}}]{Timmons2020}%
  \BibitemOpen
  \bibfield  {author} {\bibinfo {author} {\bibfnamefont {E.~I.}\ \bibnamefont
  {Timmons}}, \bibinfo {author} {\bibfnamefont {S.}~\bibnamefont
  {Teknowijoyo}}, \bibinfo {author} {\bibfnamefont {M.}~\bibnamefont
  {Ko\ifmmode~\acute{n}\else \'{n}\fi{}czykowski}}, \bibinfo {author}
  {\bibfnamefont {O.}~\bibnamefont {Cavani}}, \bibinfo {author} {\bibfnamefont
  {M.~A.}\ \bibnamefont {Tanatar}}, \bibinfo {author} {\bibfnamefont
  {S.}~\bibnamefont {Ghimire}}, \bibinfo {author} {\bibfnamefont
  {K.}~\bibnamefont {Cho}}, \bibinfo {author} {\bibfnamefont {Y.}~\bibnamefont
  {Lee}}, \bibinfo {author} {\bibfnamefont {L.}~\bibnamefont {Ke}}, \bibinfo
  {author} {\bibfnamefont {N.~H.}\ \bibnamefont {Jo}}, \bibinfo {author}
  {\bibfnamefont {S.~L.}\ \bibnamefont {Bud'ko}}, \bibinfo {author}
  {\bibfnamefont {P.~C.}\ \bibnamefont {Canfield}}, \bibinfo {author}
  {\bibfnamefont {P.~P.}\ \bibnamefont {Orth}}, \bibinfo {author}
  {\bibfnamefont {M.~S.}\ \bibnamefont {Scheurer}}, \ and\ \bibinfo {author}
  {\bibfnamefont {R.}~\bibnamefont {Prozorov}},\ }\bibfield  {title} {\enquote
  {\bibinfo {title} {{Electron irradiation effects on superconductivity in
  ${\mathrm{PdTe}}_{2}$: An application of a generalized Anderson theorem}},}\
  }\href {\doibase 10.1103/PhysRevResearch.2.023140} {\bibfield  {journal}
  {\bibinfo  {journal} {Phys. Rev. Research}\ }\textbf {\bibinfo {volume}
  {2}},\ \bibinfo {pages} {023140} (\bibinfo {year} {2020})}\BibitemShut
  {NoStop}%
\bibitem [{\citenamefont {Nakosai}\ \emph {et~al.}(2012)\citenamefont
  {Nakosai}, \citenamefont {Tanaka},\ and\ \citenamefont
  {Nagaosa}}]{Nakosai2012}%
  \BibitemOpen
  \bibfield  {author} {\bibinfo {author} {\bibfnamefont {S.}~\bibnamefont
  {Nakosai}}, \bibinfo {author} {\bibfnamefont {Y.}~\bibnamefont {Tanaka}}, \
  and\ \bibinfo {author} {\bibfnamefont {N.}~\bibnamefont {Nagaosa}},\
  }\bibfield  {title} {\enquote {\bibinfo {title} {{Topological
  Superconductivity in Bilayer Rashba System}},}\ }\href {\doibase
  10.1103/PhysRevLett.108.147003} {\bibfield  {journal} {\bibinfo  {journal}
  {Phys. Rev. Lett.}\ }\textbf {\bibinfo {volume} {108}},\ \bibinfo {pages}
  {147003} (\bibinfo {year} {2012})}\BibitemShut {NoStop}%
\bibitem [{\citenamefont {Zhu}(2016)}]{Zhu2016}%
  \BibitemOpen
  \bibfield  {author} {\bibinfo {author} {\bibfnamefont {J.-X.}\ \bibnamefont
  {Zhu}},\ }\href {\doibase 10.1007/978-3-319-31314-6} {\emph {\bibinfo {title}
  {Bogoliubov-de Gennes Method and Its Applications}}}\ (\bibinfo  {publisher}
  {Springer International Publishing},\ \bibinfo {year} {2016})\BibitemShut
  {NoStop}%
\bibitem [{\citenamefont {Ortuzar}\ \emph {et~al.}(2022)\citenamefont
  {Ortuzar}, \citenamefont {Trivini}, \citenamefont {Alvarado}, \citenamefont
  {Rouco}, \citenamefont {Zaldivar}, \citenamefont {Yeyati}, \citenamefont
  {Pascual},\ and\ \citenamefont {Bergeret}}]{Ortuzar2022}%
  \BibitemOpen
  \bibfield  {author} {\bibinfo {author} {\bibfnamefont {J.}~\bibnamefont
  {Ortuzar}}, \bibinfo {author} {\bibfnamefont {S.}~\bibnamefont {Trivini}},
  \bibinfo {author} {\bibfnamefont {M.}~\bibnamefont {Alvarado}}, \bibinfo
  {author} {\bibfnamefont {M.}~\bibnamefont {Rouco}}, \bibinfo {author}
  {\bibfnamefont {J.}~\bibnamefont {Zaldivar}}, \bibinfo {author}
  {\bibfnamefont {A.~L.}\ \bibnamefont {Yeyati}}, \bibinfo {author}
  {\bibfnamefont {J.~I.}\ \bibnamefont {Pascual}}, \ and\ \bibinfo {author}
  {\bibfnamefont {F.~S.}\ \bibnamefont {Bergeret}},\ }\bibfield  {title}
  {\enquote {\bibinfo {title} {{Yu-Shiba-Rusinov states in two-dimensional
  superconductors with arbitrary Fermi contours}},}\ }\href {\doibase
  10.1103/PhysRevB.105.245403} {\bibfield  {journal} {\bibinfo  {journal}
  {Phys. Rev. B}\ }\textbf {\bibinfo {volume} {105}},\ \bibinfo {pages}
  {245403} (\bibinfo {year} {2022})}\BibitemShut {NoStop}%
\bibitem [{\citenamefont {Uldemolins}\ \emph {et~al.}(2022)\citenamefont
  {Uldemolins}, \citenamefont {Mesaros},\ and\ \citenamefont
  {Simon}}]{Uldemolins2022}%
  \BibitemOpen
  \bibfield  {author} {\bibinfo {author} {\bibfnamefont {M.}~\bibnamefont
  {Uldemolins}}, \bibinfo {author} {\bibfnamefont {A.}~\bibnamefont {Mesaros}},
  \ and\ \bibinfo {author} {\bibfnamefont {P.}~\bibnamefont {Simon}},\
  }\bibfield  {title} {\enquote {\bibinfo {title} {Quasiparticle focusing of
  bound states in two-dimensional $s$-wave superconductors},}\ }\href {\doibase
  10.1103/PhysRevB.105.144503} {\bibfield  {journal} {\bibinfo  {journal}
  {Phys. Rev. B}\ }\textbf {\bibinfo {volume} {105}},\ \bibinfo {pages}
  {144503} (\bibinfo {year} {2022})}\BibitemShut {NoStop}%
\bibitem [{\citenamefont {Kariyado}\ and\ \citenamefont
  {Ogata}(2010)}]{Kariyado2010}%
  \BibitemOpen
  \bibfield  {author} {\bibinfo {author} {\bibfnamefont {T.}~\bibnamefont
  {Kariyado}}\ and\ \bibinfo {author} {\bibfnamefont {M.}~\bibnamefont
  {Ogata}},\ }\bibfield  {title} {\enquote {\bibinfo {title} {{Single-Impurity
  Problem in Iron-Pnictide Superconductors}},}\ }\href {\doibase
  10.1143/JPSJ.79.083704} {\bibfield  {journal} {\bibinfo  {journal} {Journal
  of the Physical Society of Japan}\ }\textbf {\bibinfo {volume} {79}},\
  \bibinfo {pages} {083704} (\bibinfo {year} {2010})}\BibitemShut {NoStop}%
\bibitem [{\citenamefont {Beaird}\ \emph {et~al.}(2012)\citenamefont {Beaird},
  \citenamefont {Vekhter},\ and\ \citenamefont {Zhu}}]{Beaird2012}%
  \BibitemOpen
  \bibfield  {author} {\bibinfo {author} {\bibfnamefont {R.}~\bibnamefont
  {Beaird}}, \bibinfo {author} {\bibfnamefont {I.}~\bibnamefont {Vekhter}}, \
  and\ \bibinfo {author} {\bibfnamefont {J.-X.}\ \bibnamefont {Zhu}},\
  }\bibfield  {title} {\enquote {\bibinfo {title} {Impurity states in multiband
  $s$-wave superconductors: Analysis of iron pnictides},}\ }\href {\doibase
  10.1103/PhysRevB.86.140507} {\bibfield  {journal} {\bibinfo  {journal} {Phys.
  Rev. B}\ }\textbf {\bibinfo {volume} {86}},\ \bibinfo {pages} {140507}
  (\bibinfo {year} {2012})}\BibitemShut {NoStop}%
\bibitem [{\citenamefont {Yip}(2013)}]{Yip2013}%
  \BibitemOpen
  \bibfield  {author} {\bibinfo {author} {\bibfnamefont {S.-K.}\ \bibnamefont
  {Yip}},\ }\bibfield  {title} {\enquote {\bibinfo {title} {{Models of
  superconducting Cu:Bi${}_{2}$Se${}_{3}$: Single- versus two-band
  description}},}\ }\href {\doibase 10.1103/PhysRevB.87.104505} {\bibfield
  {journal} {\bibinfo  {journal} {Phys. Rev. B}\ }\textbf {\bibinfo {volume}
  {87}},\ \bibinfo {pages} {104505} (\bibinfo {year} {2013})}\BibitemShut
  {NoStop}%
\bibitem [{\citenamefont {Kobayashi}\ and\ \citenamefont
  {Sato}(2015)}]{Kobayashi2015}%
  \BibitemOpen
  \bibfield  {author} {\bibinfo {author} {\bibfnamefont {S.}~\bibnamefont
  {Kobayashi}}\ and\ \bibinfo {author} {\bibfnamefont {M.}~\bibnamefont
  {Sato}},\ }\bibfield  {title} {\enquote {\bibinfo {title} {{Topological
  Superconductivity in Dirac Semimetals}},}\ }\href {\doibase
  10.1103/PhysRevLett.115.187001} {\bibfield  {journal} {\bibinfo  {journal}
  {Phys. Rev. Lett.}\ }\textbf {\bibinfo {volume} {115}},\ \bibinfo {pages}
  {187001} (\bibinfo {year} {2015})}\BibitemShut {NoStop}%
\bibitem [{\citenamefont {Hashimoto}\ \emph {et~al.}(2016)\citenamefont
  {Hashimoto}, \citenamefont {Kobayashi}, \citenamefont {Tanaka},\ and\
  \citenamefont {Sato}}]{Hashimoto2016}%
  \BibitemOpen
  \bibfield  {author} {\bibinfo {author} {\bibfnamefont {T.}~\bibnamefont
  {Hashimoto}}, \bibinfo {author} {\bibfnamefont {S.}~\bibnamefont
  {Kobayashi}}, \bibinfo {author} {\bibfnamefont {Y.}~\bibnamefont {Tanaka}}, \
  and\ \bibinfo {author} {\bibfnamefont {M.}~\bibnamefont {Sato}},\ }\bibfield
  {title} {\enquote {\bibinfo {title} {{Superconductivity in doped Dirac
  semimetals}},}\ }\href {\doibase 10.1103/PhysRevB.94.014510} {\bibfield
  {journal} {\bibinfo  {journal} {Phys. Rev. B}\ }\textbf {\bibinfo {volume}
  {94}},\ \bibinfo {pages} {014510} (\bibinfo {year} {2016})}\BibitemShut
  {NoStop}%
\bibitem [{\citenamefont {Yanase}(2016)}]{Yanase2016}%
  \BibitemOpen
  \bibfield  {author} {\bibinfo {author} {\bibfnamefont {Y.}~\bibnamefont
  {Yanase}},\ }\bibfield  {title} {\enquote {\bibinfo {title} {Nonsymmorphic
  {W}eyl superconductivity in {UPt}$_{3}$ based on ${E}_{2u}$
  representation},}\ }\href {\doibase 10.1103/PhysRevB.94.174502} {\bibfield
  {journal} {\bibinfo  {journal} {Phys. Rev. B}\ }\textbf {\bibinfo {volume}
  {94}},\ \bibinfo {pages} {174502} (\bibinfo {year} {2016})}\BibitemShut
  {NoStop}%
\bibitem [{\citenamefont {Xie}\ \emph {et~al.}(2020)\citenamefont {Xie},
  \citenamefont {Zhou},\ and\ \citenamefont {Law}}]{Xie_2020}%
  \BibitemOpen
  \bibfield  {author} {\bibinfo {author} {\bibfnamefont {Y.-M.}\ \bibnamefont
  {Xie}}, \bibinfo {author} {\bibfnamefont {B.~T.}\ \bibnamefont {Zhou}}, \
  and\ \bibinfo {author} {\bibfnamefont {K.~T.}\ \bibnamefont {Law}},\
  }\bibfield  {title} {\enquote {\bibinfo {title} {{Spin-Orbit-Parity-Coupled
  Superconductivity in Topological Monolayer {${\mathrm{WTe}}_{2}$}}},}\ }\href
  {\doibase 10.1103/PhysRevLett.125.107001} {\bibfield  {journal} {\bibinfo
  {journal} {Phys. Rev. Lett.}\ }\textbf {\bibinfo {volume} {125}},\ \bibinfo
  {pages} {107001} (\bibinfo {year} {2020})}\BibitemShut {NoStop}%
\bibitem [{\citenamefont {Shishidou}\ \emph {et~al.}(2021)\citenamefont
  {Shishidou}, \citenamefont {Suh}, \citenamefont {Brydon}, \citenamefont
  {Weinert},\ and\ \citenamefont {Agterberg}}]{Shishidou_UTe2_2021}%
  \BibitemOpen
  \bibfield  {author} {\bibinfo {author} {\bibfnamefont {T.}~\bibnamefont
  {Shishidou}}, \bibinfo {author} {\bibfnamefont {H.~G.}\ \bibnamefont {Suh}},
  \bibinfo {author} {\bibfnamefont {P.~M.~R.}\ \bibnamefont {Brydon}}, \bibinfo
  {author} {\bibfnamefont {M.}~\bibnamefont {Weinert}}, \ and\ \bibinfo
  {author} {\bibfnamefont {D.~F.}\ \bibnamefont {Agterberg}},\ }\bibfield
  {title} {\enquote {\bibinfo {title} {{Topological band and superconductivity
  in ${\mathrm{UTe}}_{2}$}},}\ }\href {\doibase 10.1103/PhysRevB.103.104504}
  {\bibfield  {journal} {\bibinfo  {journal} {Phys. Rev. B}\ }\textbf {\bibinfo
  {volume} {103}},\ \bibinfo {pages} {104504} (\bibinfo {year}
  {2021})}\BibitemShut {NoStop}%
\bibitem [{\citenamefont {Perrin}\ \emph {et~al.}(2020)\citenamefont {Perrin},
  \citenamefont {Santos}, \citenamefont {M\'enard}, \citenamefont {Brun},
  \citenamefont {Cren}, \citenamefont {Civelli},\ and\ \citenamefont
  {Simon}}]{Perrin2020}%
  \BibitemOpen
  \bibfield  {author} {\bibinfo {author} {\bibfnamefont {V.}~\bibnamefont
  {Perrin}}, \bibinfo {author} {\bibfnamefont {F.~L.~N.}\ \bibnamefont
  {Santos}}, \bibinfo {author} {\bibfnamefont {G.~C.}\ \bibnamefont
  {M\'enard}}, \bibinfo {author} {\bibfnamefont {C.}~\bibnamefont {Brun}},
  \bibinfo {author} {\bibfnamefont {T.}~\bibnamefont {Cren}}, \bibinfo {author}
  {\bibfnamefont {M.}~\bibnamefont {Civelli}}, \ and\ \bibinfo {author}
  {\bibfnamefont {P.}~\bibnamefont {Simon}},\ }\bibfield  {title} {\enquote
  {\bibinfo {title} {{Unveiling Odd-Frequency Pairing around a Magnetic
  Impurity in a Superconductor}},}\ }\href {\doibase
  10.1103/PhysRevLett.125.117003} {\bibfield  {journal} {\bibinfo  {journal}
  {Phys. Rev. Lett.}\ }\textbf {\bibinfo {volume} {125}},\ \bibinfo {pages}
  {117003} (\bibinfo {year} {2020})}\BibitemShut {NoStop}%
\bibitem [{\citenamefont {Suzuki}\ \emph {et~al.}(2022)\citenamefont {Suzuki},
  \citenamefont {Sato},\ and\ \citenamefont {Asano}}]{Suzuki2022}%
  \BibitemOpen
  \bibfield  {author} {\bibinfo {author} {\bibfnamefont {S.-I.}\ \bibnamefont
  {Suzuki}}, \bibinfo {author} {\bibfnamefont {T.}~\bibnamefont {Sato}}, \ and\
  \bibinfo {author} {\bibfnamefont {Y.}~\bibnamefont {Asano}},\ }\bibfield
  {title} {\enquote {\bibinfo {title} {{Odd-frequency Cooper pair around a
  magnetic impurity}},}\ }\href {\doibase 10.1103/PhysRevB.106.104518}
  {\bibfield  {journal} {\bibinfo  {journal} {Phys. Rev. B}\ }\textbf {\bibinfo
  {volume} {106}},\ \bibinfo {pages} {104518} (\bibinfo {year}
  {2022})}\BibitemShut {NoStop}%
\bibitem [{\citenamefont {Triola}\ \emph {et~al.}(2020)\citenamefont {Triola},
  \citenamefont {Cayao},\ and\ \citenamefont {Black-Schaffer}}]{Triola2020}%
  \BibitemOpen
  \bibfield  {author} {\bibinfo {author} {\bibfnamefont {C.}~\bibnamefont
  {Triola}}, \bibinfo {author} {\bibfnamefont {J.}~\bibnamefont {Cayao}}, \
  and\ \bibinfo {author} {\bibfnamefont {A.~M.}\ \bibnamefont
  {Black-Schaffer}},\ }\bibfield  {title} {\enquote {\bibinfo {title} {{The
  Role of Odd-Frequency Pairing in Multiband Superconductors}},}\ }\href
  {\doibase https://doi.org/10.1002/andp.201900298} {\bibfield  {journal}
  {\bibinfo  {journal} {Annalen der Physik}\ }\textbf {\bibinfo {volume}
  {532}},\ \bibinfo {pages} {1900298} (\bibinfo {year} {2020})}\BibitemShut
  {NoStop}%
\end{thebibliography}%

\appendix

\section{\label{sec:Appendix Green's function}Momentum-averaged
  Green's function}

The Green's function is formally written as
\begin{equation}
G_0({\bf k},i\omega_n) = (i\omega_n - {\cal H}_{\text{BdG}}({\bm
  k}))^{-1} \label{eq:G0formal}
\end{equation}
where ${\cal H}_{\text{BdG}}({\bm k})$ is the BdG
Hamiltonian
\begin{equation}
{\cal H}_{\text{BdG}}({\bm k}) = \begin{pmatrix}
H_{0,{\bm k}} & \Delta \\
\Delta^\dagger & -H^T_{0,{\bm k}}
\end{pmatrix}
\end{equation}
where $\Delta$ is the pairing potential and $H_{0,{\bm k}}$ is the
normal-state Hamiltonian. The two-dimensional and three-dimensional
models are particular examples of a normal-state Hamiltonian with the
generic form
\begin{equation}
H_{0,{\bm k}} = \epsilon_0\hat{\mathbf{1}}_{4}  + \vec{\epsilon}\cdot\vec{\gamma}
\end{equation}
where $\vec{\gamma} = (\gamma_1,\gamma_2,\gamma_3,\gamma_4,\gamma_5)$
are the Euclidean Dirac matrices and $\vec{\epsilon} =
                  (\epsilon_1,\epsilon_2,\epsilon_3,\epsilon_4,\epsilon_5)$
                  is the vector of the corresponding coefficients. The
                  normal-state Hamiltonian describes a two-band system
                  with energies
\begin{equation}
\epsilon_{\pm,{\bm k}} = \epsilon_0 \pm |\vec{\epsilon}|\,.
\end{equation}
In the context of our bilayer model the $\gamma$-matrices are given by
\begin{equation}
\vec{\gamma} = (\eta_x\sigma_0,\eta_y\sigma_0,\eta_z\sigma_x,\eta_z\sigma_y,\eta_z\sigma_z)
\end{equation}
with coefficients
\begin{align}
\epsilon_0=&\begin{cases}
-2t(\cos(k_xa)+\cos(k_ya))-\mu & \text{2D}\\
-\mu & \text{3D}
\end{cases}\\
\vec{\epsilon}=&\begin{cases}
(t_\perp,0,\alpha\sin(k_ya),-\alpha\sin(k_xa),0) & \text{2D}\\
(m,v_zk_z,vk_y,-vk_x,0) & \text{3D}
\end{cases}
\end{align}
In the following we will express all results in the general notation. 

\subsection{Full Green's functions}

Performing the matrix inverse in Eq.~\ref{eq:G0formal} we obtain the
full Green's functions for the 
$A_{\text{1g}}$, $A_{\text{1u}}$, and $A_{\text{2u}}$ pairing states. Neglecting terms
which average to zero across the Fermi surface we have
\begin{widetext}
\begin{equation}
\begin{aligned}[b]
G_{0}({\bm k},i\omega_n) =&
\frac{1}{\left(\omega_{n}^{2}+E_{-}^{2}\right)\left(\omega_{n}^{2}+E_{+}^{2}\right)}\left\{-i
  \omega_{n}\left(\omega_{n}^{2}+\varepsilon_{0}^{2}+|\vec{\varepsilon}|^{2}+\Delta_{0}^{2}\right)
  \hat{\mathbf{1}}_{4}  \tau_{0}  -\varepsilon_{0}\left(\omega_{n}^{2}+\varepsilon_{0}^{2}-|\vec{\varepsilon}|^{2}+\Delta_{0}^{2}\right)
  \hat{\mathbf{1}}_{4}  \tau_{z} \right.\\
& \left. +2 i \omega_{n} \varepsilon_{0} \varepsilon_{1} \gamma_{1}  \tau_{0}-\varepsilon_{1}\left(\omega_{n}^{2}-\varepsilon_{0}^{2}+|\varepsilon|^{2}+\Delta_{0}^{2}\right) \gamma_{1}  \tau_{z} + F\right\},
\end{aligned}\label{intG}
\end{equation}
where the anomalous component $F$ is 
\begin{equation}
F = \begin{cases}
-2 \Delta_{0} \varepsilon_{0}\varepsilon_{1} i\gamma_{2}\gamma_{4}  \tau_{x}
-\Delta_{0}\left(\omega_{n}^{2}+\varepsilon_{0}^{2}+|\vec{\varepsilon}|^{2}+\Delta_{0}^{2}\right)
i \gamma_{3} \gamma_{5}  \tau_{x}
& A_{\text{1g}}, \quad \Delta=\Delta_0\eta_0 i\sigma_y = \Delta_0i\gamma_3\gamma_5 \\
-2 i \omega_{n} \Delta_{0} \varepsilon_{1} \gamma_{3}  \tau_{y}
-\Delta_{0}\left(\omega_{n}^{2}+\varepsilon_{0}^{2}-2
    \varepsilon_{1}^{2}-2
    \varepsilon_{5}^{2}+|\vec{\varepsilon}|^{2}+\Delta_{0}^{2}\right)
  i \gamma_{1} \gamma_{3}  \tau_{x} & A_{\text{1u}}, \quad
  \Delta=\Delta_0\eta_y \sigma_zi\sigma_y =
  \Delta_0i\gamma_1\gamma_3\\
-2 i \omega_{n} \Delta_{0} \varepsilon_{1} i \gamma_{1}\gamma_{4}
 \tau_{x}-\Delta_{0}\left(\omega_{n}^{2}+\varepsilon_{0}^{2}-2
  \varepsilon_{1}^{2}-2
  \varepsilon_{2}^{2}+|\vec{\varepsilon}|^{2}+\Delta_{0}^{2}\right)
\gamma_{4}  \tau_{y} & A_{\text{2u}}, \quad\Delta=\Delta_0\eta_z i\sigma_y = \Delta_0i\gamma_4
\end{cases}
\end{equation}
\end{widetext}
and the dispersion $E_{\pm}$ is given by
\begin{equation}
E_{\pm} = \begin{cases}
\sqrt{(|\epsilon_0|\pm|\vec{\epsilon}|)^2 + \Delta_0^2} & A_{\text{1g}}\\
 \sqrt{\epsilon_0^2+|\vec{\epsilon}|^2 +\Delta_0^2\pm
  2\sqrt{\epsilon_0^2|\vec{\epsilon}|^2 +
    \Delta_0^2(\epsilon_1^2+\epsilon_5^2)}} & A_{\text{1u}}\\
 \sqrt{\epsilon_0^2+|\vec{\epsilon}|^2 +\Delta_0^2\pm
  2\sqrt{\epsilon_0^2|\vec{\epsilon}|^2 +
    \Delta_0^2(\epsilon_1^2+\epsilon_2^2)}} & A_{\text{2u}}\\
\end{cases}
\end{equation}
The Fermi surface of the noninteracting system is defined by
$|\epsilon_0|-|\vec{\epsilon}|=0$. This is equivalent to
$\lim_{\Delta_0\rightarrow0}E_{-}=0$, and so $E_-$ corresponds to the
low-energy dispersion. For the odd-parity states and
$|\Delta_0|\ll|\vec{\epsilon}|$, we can approximate $E_-$ in this region by
\begin{equation}
E_{-} \approx 
 \sqrt{(|\epsilon_0|-|\vec{\epsilon}|)^2 +\Delta_{\text{eff}}^2} 
\end{equation}
where $\Delta_{\text{eff}} = \Delta_0\sqrt{1-\tilde{F}_\nu}$ is the
effective gap introduced in Eq.~\ref{eq:effgap}.

\subsection{Low-energy approximation}

Our aim here is to derive an expression for the Green's function which
is valid close to the Fermi surface. 
Formally we can decompose the Green's function as
\begin{equation}
G_0 = \frac{G_0^{+}}{2(\omega_n^2+E_{+}^2)} + \frac{G_0^{-}}{2(\omega_n^2+E_{-}^2)}
\end{equation}
Since the $-$ branch corresponds to the low-energy dispersion,
discarding the contribution from the $+$ branch gives us the Green's
function projected onto a low-energy subspace. Neglecting terms of order
$\Delta_0/|\vec{\epsilon}|$ and higher, we obtain
\begin{align}
G^{-}_{0} \approx& -i\omega_n  \hat{\mathbf{1}}_{4} 
\tau_{0} + \text{sgn}(\epsilon_0)\xi_- \hat{\mathbf{1}}_{4} 
\tau_{z} \notag \\
& +i\omega_n\text{sgn}(\epsilon_0)\hat{\epsilon}_1\gamma_1\tau_0
- \hat{\epsilon}_1\xi_-\gamma_1\tau_z \notag\\
& +\begin{cases}
-\text{sgn}(\epsilon_0)\hat{\epsilon}_1\Delta_0i\gamma_2\gamma_4\tau_x
- \Delta_0i\gamma_3\gamma_5\tau_x & A_{\text{1g}}\\
-\Delta_0(1 -
\hat{\epsilon}_1^2-\hat{\epsilon}_5^2)i\gamma_1\gamma_3\tau_x &
A_{\text{1u}}\\
-\Delta_0(1 -
\hat{\epsilon}_1^2-\hat{\epsilon}_2^2)\gamma_4\tau_y &
A_{\text{2u}}
\end{cases}
\end{align}
Note that the
Green's function depends on the sign of $\epsilon_0$. Fermi surfaces
where $\epsilon_0$ have different signs corresponding to different
bands. In the
cases we consider we have $\epsilon_0=-\mu<0$, but our analysis
remains valid for a momentum-dependent $\epsilon_0$ so long as only
one band crosses the Fermi energy.

To finally obtain the momentum-averaged Green's function Eq.~\ref{eq:Greensfunction} we
perform the integral over $\xi_-$ and an average over the Fermi
surface as shown in Eq.~\ref{eq:sumtoint}. With the exception of the
coefficients of $\hat{\mathbf{1}}_{4} \tau_z$ and
$\gamma_1\tau_z$, these integrals converge as
$\Lambda\rightarrow\infty$ and so we approximate
\begin{align}
\int_{-\Lambda}^{\Lambda} 
\frac{N(\xi_{-})}{\omega_{n}^{2}+\xi_{-}^{2}+\Delta_{\text{eff}}^{2}} d
\xi_{-}
\approx &\int_{-\infty}^{\infty} 
\frac{N(\xi_{-})}{\omega_{n}^{2}+\xi_{-}^{2}+\Delta_{\text{eff}}^{2}} d
\xi_{-} \notag \\
=  &\frac{N_0 \pi}{\sqrt{\omega_{n}^{2}+\Delta_{\text{eff}}^{2}}}
\end{align}
The coefficients of $\hat{\mathbf{1}}_{4} \tau_z$ and
$\gamma_1\tau_z$ are only nonzero when the particle-hole
asymmetry of the normal-state DOS is taken into account. These terms
are cutoff-dependent, and in the limit $\Lambda \gg
|\Delta_{\text{eff}}|$ we can approximate them as
\begin{equation}
\int_{-\Lambda}^{\Lambda} 
\frac{N(\xi_{-})\xi_-}{\omega_{n}^{2}+\xi_{-}^{2}+\Delta_{\text{eff}}^{2}} d
\xi_{-} \approx - 2N_{0}^{\prime} \Lambda \equiv-2y, 
\end{equation}
and we define $\tilde{y}= y/(\frac{\pi}{2}N_{0})$ as a dimensionless fitting parameter.

\section{\label{sec:AppendixAnalytic}Analytical expression for bound
  states with particle-hole asymmetry.}

It is necessary to include the particle-hole asymmetry parameter $y$
to obtain quantitative agreement between the exact diagonalization and
the $T$-matrix results for the impurity bound states. These
expressions are complicated and are given here for completeness.

\subsection{\label{sec:Appendix A1g BS}$A_{\text{1g}}$ pairing}

Impurity bound states are realized in the class $a^\ast$ and $b^\ast$,
and have a general dependence on the normalized impurity potential
strength 
\begin{equation}
\tilde{\omega}= \pm
\frac{1-\beta_2 \tilde{V}^2}
{\sqrt{1+\alpha_2 \tilde{V}^2 +\alpha_4 \tilde{V}^4}}
\end{equation}
The coefficients $\beta_i$, $\alpha_j$ are given by
\begin{itemize}
\item class $a^{*}$: $\beta_2= (1-\tilde{m}^2)(1 + \tilde{y}^2)$, $\alpha_2=2(1 -
  \tilde{m}^2)(1- \tilde{y}^2)$ and $\alpha_4=((1 - \tilde{m}^2)(1+ \tilde{y}^2))^2$. 
\item class $b^{*}$: $\beta_2^{\pm}= (1\pm \tilde{m})^2(1 + \tilde{y}^2)$, $\alpha_2^{\pm}=2(1\pm \tilde{m})^2(1- \tilde{y}^2)$ and $\alpha_4^{\pm}=((1\pm \tilde{m})^2(1+ \tilde{y}^2))^2$.
\end{itemize}

\subsection{\label{sec:Appendix six BS}$A_{\text{1u}}$ and $A_{\text{2u}}$ pairing.}

Impurity bound states are realized in all classes except $c$, and have
the general dependence on the normalized impurity potential 
strength
\begin{equation}
\tilde{\omega}= \pm
\frac{1+\beta_1 \tilde{V}+\beta_2 \tilde{V}^2}
{\sqrt{1+\alpha_1 \tilde{V} +\alpha_2 \tilde{V}^2 +\alpha_3 \tilde{V}^3 +\alpha_4 \tilde{V}^4}}
\end{equation}
where the coefficients $\beta_i$ and $\alpha_j$ are given 
by
\begin{itemize}
\item class $a$: $\beta_1=2 \tilde{y}$, $\beta_2= ( 1-\tilde{m}^2)(1 + \tilde{y}^2)$, $\alpha_1=4 \tilde{y}$, $\alpha_2=(2 + 2 \tilde{m}^2 + (6 -2 \tilde{m}^2) \tilde{y}^2)$,
$\alpha_3= 4\tilde{y}(1-\tilde{m}^2)(1+ \tilde{y}^2)$ and $\alpha_4=((1 - \tilde{m}^2)(1+ \tilde{y}^2))^2$.
\item class $b$:
$\beta_1=2\tilde{m} \tilde{y}$, $\beta_2= ( \tilde{m}^2-1)(1 + \tilde{y}^2)$, $\alpha_1=4 \tilde{m}\tilde{y}$, $\alpha_2=(2 + 2 \tilde{m}^2 + (-2 +6 \tilde{m}^2) \tilde{y}^2)$,
$\alpha_3= 4\tilde{m}\tilde{y}(\tilde{m}^2-1)(1+ \tilde{y}^2)$ and $\alpha_4=((\tilde{m}^2-1)(1+ \tilde{y}^2))^2$.  
\item class $d$:
  $\beta_1=0$, $\beta_2= (\tilde{m}^2-1)(1 + \tilde{y}^2)$, $\alpha_1=0$, $\alpha_2=-2( \tilde{m}^2-1)(1- \tilde{y}^2)$,
$\alpha_3= 0$ and $\alpha_4=(( \tilde{m}^2-1)(1+ \tilde{y}^2))^2$.
\item class $e$: 
$\beta_1^{\pm}=\pm 2 \tilde{y}$, $\beta_2^{\pm}= ( 1-\tilde{m}^2)(1 + \tilde{y}^2)$,
$\alpha_1^{\pm}=\pm 4 \tilde{y}$, $\alpha_2^{\pm}=(2 + 2 \tilde{m}^2 + (6 - 2 \tilde{m}^2) \tilde{y}^2)$,
 $\alpha_3^{\pm}=\pm 4\tilde{y}(1-\tilde{m}^2)(1+ \tilde{y}^2)$ and $\alpha_4^{\pm}=((1 - \tilde{m}^2)(1+ \tilde{y}^2))^2$.  
\item class $f$: $\beta_1^{\pm}=\pm 2\tilde{m} \tilde{y}$, $\beta_2^{\pm}= ( \tilde{m}^2-1)(1
  + \tilde{y}^2)$, $\alpha_1^{\pm}=\pm 4\tilde{m} \tilde{y}$, $\alpha_2^{\pm}=(2 + 2 \tilde{m}^2 + (-2 + 6 \tilde{m}^2) \tilde{y}^2)$,
$\alpha_3^{\pm}= \pm 4\tilde{m}\tilde{y}(\tilde{m}^2-1)(1+ \tilde{y}^2)$ and $\alpha_4^{\pm}=((\tilde{m}^2-1)(1+ \tilde{y}^2))^2$.

\end{itemize}

\end{document}